# Optical library of $Ga_2O_3$ polymorphs

*Augustinas Galeckas,* Adrian Cernescu, Anna Kaźmierczak-Bałata, Javier García-Fernández, Calliope Bazioti, Alexander Azarov, Junlei Zhao, Ji-Hyeon Park, Dae-Woo Jeon, Halin Lee, Won-Jae Lee, Ray-Hua Horng, Rui Zhu, Zengxia Mei, Øystein Prytz, and Andrej Kuznetsov**

**Gallium oxide is an emerging material of interest due to its unique combination of functional properties and the existence of multiple polymorphs - *α, β, γ, δ*, and *κ* - each exhibiting distinct characteristics arising from their different lattice symmetries. Optical properties are particularly important, as they determine potential device applications and enable phase identification. However, direct comparison of optical signatures, including key parameters such as bandgaps, is hindered by inconsistent, sparse, or even missing data in the literature. To address this issue, in the present work we systematically cross-correlate optical emission and absorption features of *α, β, γ, δ* and *κ* thin films, as well as differently oriented *β*-phase bulk crystals and *γ/β* double polymorph structures. We demonstrate that optical bandgaps and emission features scale consistently across the polymorphs when methodological uncertainties are minimized by applying identical experimental conditions and unified analysis procedures to a structurally similar set of thin-film samples. In addition, we extend conventional far-field optical phase identification to the nanoscale by reporting near-field optical signatures of $Ga_2O_3$ polymorphs via nano-FTIR. Overall, the present dataset provides a comprehensive reference of near- and far-field optical polymorph signatures to support ongoing multidisciplinary research on $Ga_2O_3$.**

## 1. Introduction

Gallium oxide ($Ga_2O_3$) is an ultra-wide bandgap semiconductor with great potential for power electronics and UV photonic applications.[1,2] The technological importance of $Ga_2O_3$ is further augmented by the existence of several polymorphs, possessing monoclinic ($\beta$), rhombohedral ($\alpha$), cubic defect spinel ($\gamma$), cubic bixbyite ($\delta$), and orthorhombic ($\kappa$) crystal structure, each with distinct characteristics that favor specific device applications. For instance, the higher crystal symmetry $\alpha$- and $\kappa$-$Ga_2O_3$ phases are expected to enhance the performance of solar-blind UV optoelectronics;[3] also promising are $\alpha$-$Ga_2O_3$-based Schottky barrier diodes and metal-oxide-semiconductor field-effect transistors.[4] Moreover, the spontaneous polarization,[5,6,7] inherent to $\kappa$-$Ga_2O_3$ phase due to the lack of inversion symmetry, provides an opportunity for creating a two-dimensional electron gas (2DEG) and realization of high-performance devices that rely on the 2DEG confinement.[8,9] The monoclinic $\beta$-$Ga_2O_3$ presently dominates over other polymorphs in a wide range of device applications[2,10,11,12,13] owing to its stability and well established large size single crystal wafer fabrication. Recently, the functionality of $\beta$-$Ga_2O_3$ has been further extended through ion irradiation-induced phase transformations.[14,15,16] The $\gamma$-$Ga_2O_3$ polymorph, formed through disorder-induced ordering, demonstrates remarkably high radiation tolerance[17] and offers an opportunity to utilize unique properties of the polymorphic $\gamma/\beta$ heterostructure,[18,19] thus broadening the range

A. Galeckas, J. García-Fernández, C. Bazioti, A. Azarov, Ø. Prytz, A. Kuznetsov
Department of Physics
Centre for Materials Science and Nanotechnology
University of Oslo
PO Box 1048 Blindern, N-0316 Oslo, Norway
E-mail: augustinas.galeckas@fys.uio.no,
E-mail: andrej.kuznetsov@fys.uio.no

A. Cernescu
attocube systems AG
85540 Haar, Germany

Anna Kaźmierczak-Bałata
Institute of Physics
Silesian University of Technology
44-100 Gliwice, Poland

Junlei Zhao
The Hong Kong Microelectronics Research and Development Institute
Hong Kong, 999077, China

Ji-H. Park, D-W. Jeon, H. Lee
Korea Institute of Ceramic Engineering & Technology
Jinju 52851, Republic of Korea

W-J. Lee
Department of Advanced Materials Engineering
Dong-Eui University
47340 Busan, Republic of Korea

Ray-Hua Horng
Institute of Electronics,
National Yang Ming Chiao Tung University
Hsinchu 30010, Taiwan

R. Zhu, Z. Mei
Songshan Lake Materials Laboratory
523808 Dongguan, Guangdong, P. R. China
Institute of Physics
Chinese Academy of Sciences
100190 Beijing, P. R. China

of applications beyond what single-phase materials can achieve. In addition to the diverse applications of crystalline polymorphs, the amorphous state of $Ga_2O_3$ demonstrates significant potential as well, especially in solar-blind optoelectronics[20] and data storage technologies.[21,22]

The current understanding of the physical properties of various polymorphs varies widely. The most comprehensive studies have been conducted on the *β* polymorph, whereas the *γ* phase remains the least explored. This trend also applies to optical emission and absorption properties, including optical bandgaps, which are often subjects of controversy in the literature. Indeed, the reported data for fundamental energy gap ($E_g$) of $Ga_2O_3$, considering all polymorphs (*α*, *β*, *γ*, *δ* and *κ*), covers the range from 4.4 to 5.6 eV.[3,23,24,25,26,27] Meanwhile, the bandgap of monoclinic *β*-$Ga_2O_3$, determined from the onset of band-edge absorption, varies nearly as much, ranging from 4.4 to 5.0 eV.[28] The wide bandgap variation reported for the same polymorph is due to a combination of several methodological factors and is only partly attributable to the anisotropic crystal structure, as detailed below. Indeed, literature data for corundum-like *α*-$Ga_2O_3$, which has the widest bandgap among all polymorphs, shows a similarly wide spread of $E_g$ values (5.0 - 5.6 eV).[29,30] The optical data for the *γ*-, *δ*- and *κ*-$Ga_2O_3$ polymorphs are less extensively documented compared to other phases, with reported bandgap energies ranging from 4.6 eV to 4.9 eV for *κ*-$Ga_2O_3$,[31,32,33] from 4.8 to 5.8 eV for *δ*-$Ga_2O_3$,[34,35,36] and from 5.0 to 5.17eV for *γ*-$Ga_2O_3$.[37,38] The existing uncertainties, even within individual phases, motivate a systematic investigation of the optical absorption and emission signatures of the different $Ga_2O_3$ polymorphs, with the aim of establishing a reliable reference data repository, or optical library, which is the focus of this work.

The bandgap is commonly determined from the optical absorption edge, which, in non-cubic crystals, depends on the crystallographic direction, resulting in anisotropy. The absorption measurements typically utilize non-polarized light at normal incidence, meaning that the electric field vector $\boldsymbol{E}$ of the light lies within the surface plane. For (010) oriented *β*-$Ga_2O_3$, this means that $\boldsymbol{E}$ is perpendicular to the major crystallographic axis $\boldsymbol{b}$, and thus the anisotropy is negligible, as confirmed experimentally by polarization-dependent variation of bandgap $\Delta E_g \sim 0.03$ eV.[39] By contrast, (-201) oriented *β*-$Ga_2O_3$ with the axis $\boldsymbol{b}$ = [010] in the surface plane exhibits strong anisotropy, as evidenced by $\Delta E_g \sim 0.2$ eV blue-shift for the light polarized along the $\boldsymbol{b}$ direction.[39] In transmittance measurements using non-polarized light, the absorption edge includes both higher- and lower-energy contributions. However, the lower-energy contribution controls the onset of the absorption edge that defines the optical bandgap, thereby ruling out anisotropy as the main cause of scattered $E_g$ values in the literature for monoclinic *β*-$Ga_2O_3$. Regarding the anisotropy of other polymorphs, *α*-$Ga_2O_3$ is typically grown as a film on c-plane sapphire, following its symmetry. Therefore, polarization-dependent absorption is not expected. The same argument holds for *γ*-$Ga_2O_3$ due to its cubic crystal symmetry. Clearly, the widely spread bandgap values reported for different polymorphs cannot be attributed solely to anisotropy, but rather to a combination of anisotropy and a number of methodological factors. These include the variety of spectroscopic techniques used for bandgap assessment, such as ellipsometry, transmittance, reflectance and photoluminescence excitation (PLE), [30,40,28] as well as the different analytical procedures employed, like derivative analysis, simulation, linear regression and Tauc method [$\alpha$, $\alpha h\nu$, $(\alpha h\nu)^{1/2}$, $(\alpha h\nu)^2$].[41] The variations in structure (bulk crystals *vs*. thin films) and crystalline quality of the tested materials are also among the critical factors.

It is noteworthy that, in addition to conventional transmittance approach, several well-established methods and techniques are capable of effective polymorph characterization, each with its own advantages, limitations and practical challenges. For instance, ellipsometry is more surface-sensitive than other techniques and represents an indirect approach, requiring optical-medium modeling and extraction of $E_g$ from the dispersion of the complex refractive index. A common direct method for bandgap assessment is photoluminescence excitation (PLE), which relies on monitoring the emitted light while varying the photoexcitation wavelength. Because the photoexcitation depth decreases with increasing photon energy, nonradiative surface recombination becomes the predominant pathway at shallow excitation depths. This effect can, in turn, distort the apparent absorption edge and thereby influence the estimated $E_g$ value. On the other hand, bandgap assessment using most common transmittance method imposes certain requirements for homogeneity, thickness, and surface smoothness of the probed medium, in effect limiting its application to thin films on transparent substrates.

In this study, transmittance spectroscopy is the method of choice to investigate polymorph-related variations of the absorption edge and to assess the optical bandgaps ($E_g$) of polymorphs represented by thin films (<1 µm) grown on sapphire (c-$Al_2O_3$). Furthermore, to illustrate the systematic discrepancies in bandgap parameters due to the structure of the tested materials, bulk (thick) crystals were characterized and compared with the data for thin films. In parallel, a complementary method of diffuse-reflectance spectroscopy (DRS), which can probe both thin films on homo- and hetero-substrates as well as bulk materials, was applied for a comparative evaluation of the bandgap parameters. These experimentally derived parameters were subsequently compared with predictions obtained from density functional theory (DFT) calculations. In addition, the emission properties of the polymorphs were assessed using photoluminescence (PL) spectroscopy. Finally, we extend far-field optical phase identification to the nanoscale by gaining insight into the near-field optical signatures of the polymorphs using scanning nanoscale Fourier Transform Infrared (nano-FTIR) spectroscopy.

Importantly, optical absorption and emission properties of different polymorphs were characterized under identical, fixed experimental conditions and parameter-extraction protocols. This approach provides reliable insight into polymorph-specific features by minimizing uncertainties arising from the variety of bandgap assessment methods and techniques reported in the literature. In particular, we systematically cross-correlate the absorption edge and photoluminescence properties of $Ga_2O_3$ polymorphs, including corundum-like *α*, monoclinic *β*, cubic defect spinel *γ*, cubic bixbyite *δ*, and orthorhombic *κ*, as well as amorphous a-$Ga_2O_x$. By establishing a comparative library of near- and far-field optical signatures for various $Ga_2O_3$ polymorphs, this work aims to provide a reference framework for phase identification and for recognizing polymorph-specific optical properties.

## 2. Results and Discussions

### 2.1. Structural properties

Figure 1 presents the X-ray diffraction (XRD) 2Θ scans with indexed reflections for the $\alpha$-, $\beta$-, $\gamma$-, $\delta$-, and $\kappa$- $Ga_2O_3$ thin films grown on sapphire (panels a–c), along with the corresponding data acquired from a (010) $\beta$-$Ga_2O_3$ wafer and a $\gamma/\beta$ double-polymorph structure shown in panel (d). The displayed XRD patterns validate both the crystalline quality and nominal phase purity of the samples representing the different $Ga_2O_3$ polymorphs. Notably, the specimen representing the $\delta$-phase exhibits mixed $\delta/\kappa$ character rather than a purely single-phase structure. The uniform single-phase nature of the samples throughout the full thickness of the layers is further confirmed by TEM analysis, as evidenced by the selected-area electron diffraction (SAED) patterns and cross-sectional images of the samples in Figure 2. Importantly, the phase identification presented in Figs. 1 and 2 aligns with previously reported findings in the literature, confirming that the selected samples collectively constitute a representative set suitable for the primary objective of this study – namely, establishing an optical library of $Ga_2O_3$ polymorphs.

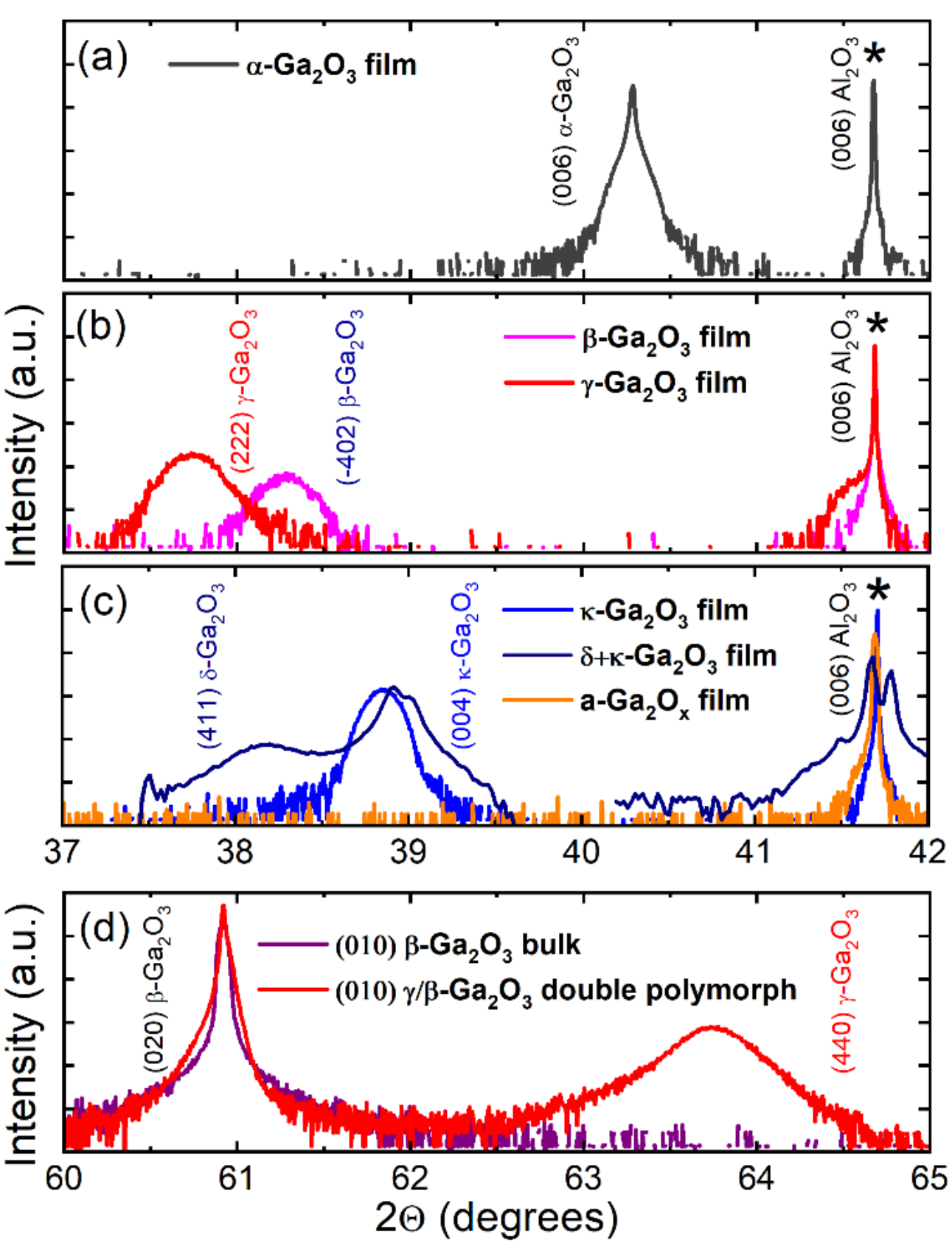

**Figure 1.** X-ray diffraction (XRD) spectra of the $\alpha$-, $\beta$-, $\gamma$-, $\delta$-, and $\kappa$-$Ga_2O_3$ polymorphs, as well as amorphous a-$Ga_2O_x$, examined in this study. Note that the low-intensity signal from the amorphous a-$Ga_2O_x$ phase in panel (c) appears primarily as background.

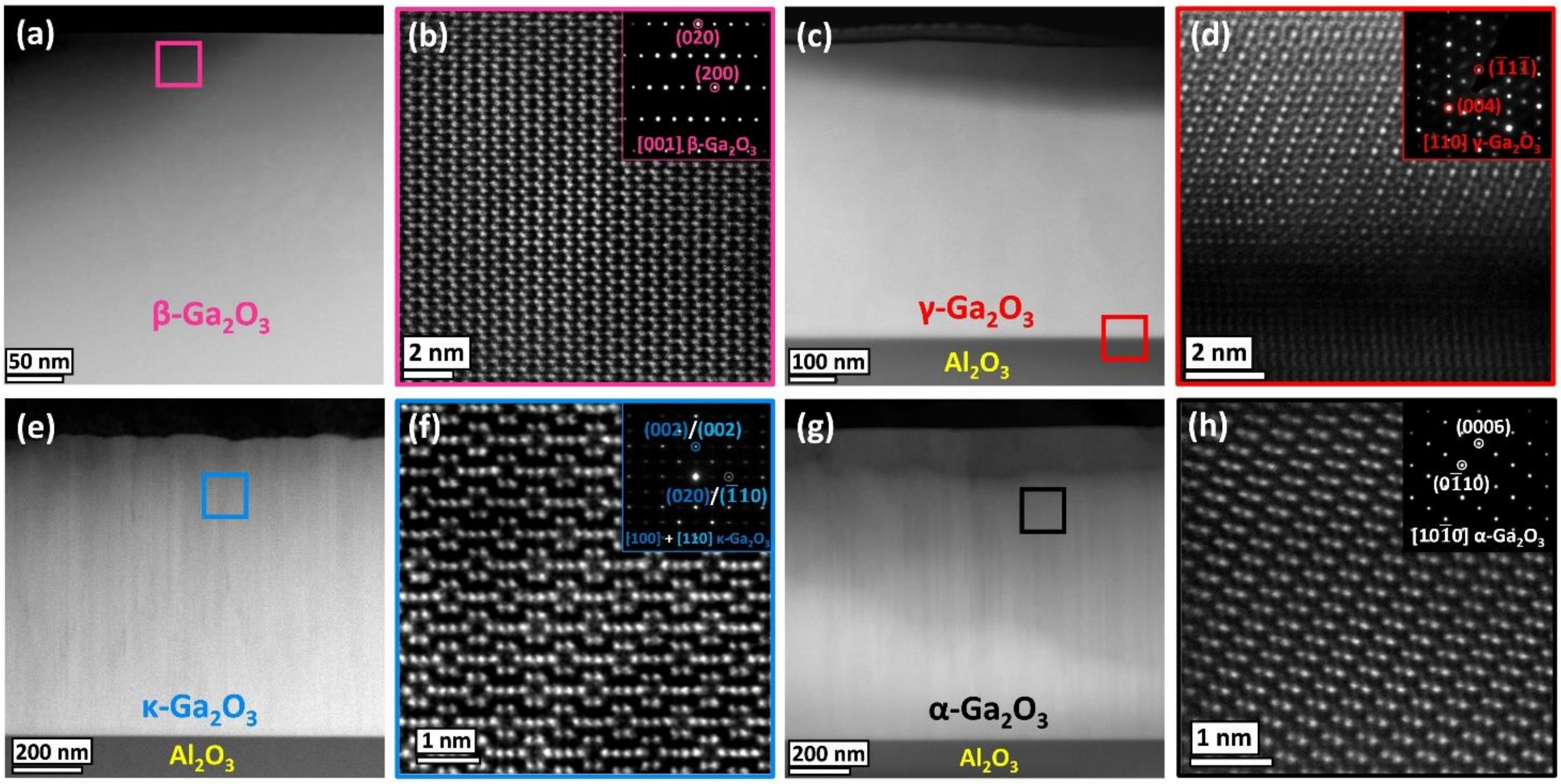

**Figure 2**. Low-magnification and corresponding atomic-resolution HAADF-STEM images of the $\beta$ -$Ga_2O_3$ (a-b), $\gamma$-$Ga_2O_3$ (c-d), $\kappa$-$Ga_2O_3$ (e-f) and $\alpha$-$Ga_2O_3$ (g-h) polymorphs. Colored boxes in the low-magnification images indicate the regions from which the high-resolution images were acquired. The insets show the SAED patterns recorded along the corresponding zone axes.

### 2.2. Absorption properties

The uniform single-phase $\alpha$-, $\beta$-, $\delta$-, and $\kappa$-$Ga_2O_3$ films grown on sapphire, as well as bulk $\beta$-$Ga_2O_3$ crystals, permit both transmittance and diffuse-reflectance measurements for bandgap assessment. Such a twofold approach presents an opportunity to interrelate the results of the two spectroscopic techniques, which are equally capable of characterizing the optical absorption edge, yet differ essentially in terms of probing depth. While the transmission of light invariably probes the entire medium, the diffuse-reflectance signal is largely determined by evanescent light penetration and light scattering at surface irregularities, which together lead to a higher sensitivity to near-surface properties than to bulk properties. The latter fact offers great advantages for studying multilayer structures, including γ thin films in double polymorph $\gamma/\beta$ structures, since the diffuse-reflectance signal can be collected in a controlled manner from the near-surface region by adjusting focal plane position within the sample. In what follows, we compare the results obtained by the two spectroscopy techniques for the different polymorph thin films grown on sapphire, comprising the primary sample set. Additional results, including measurements of bulk samples with different orientations (auxiliary subset), are provided in the Supplemental Material.

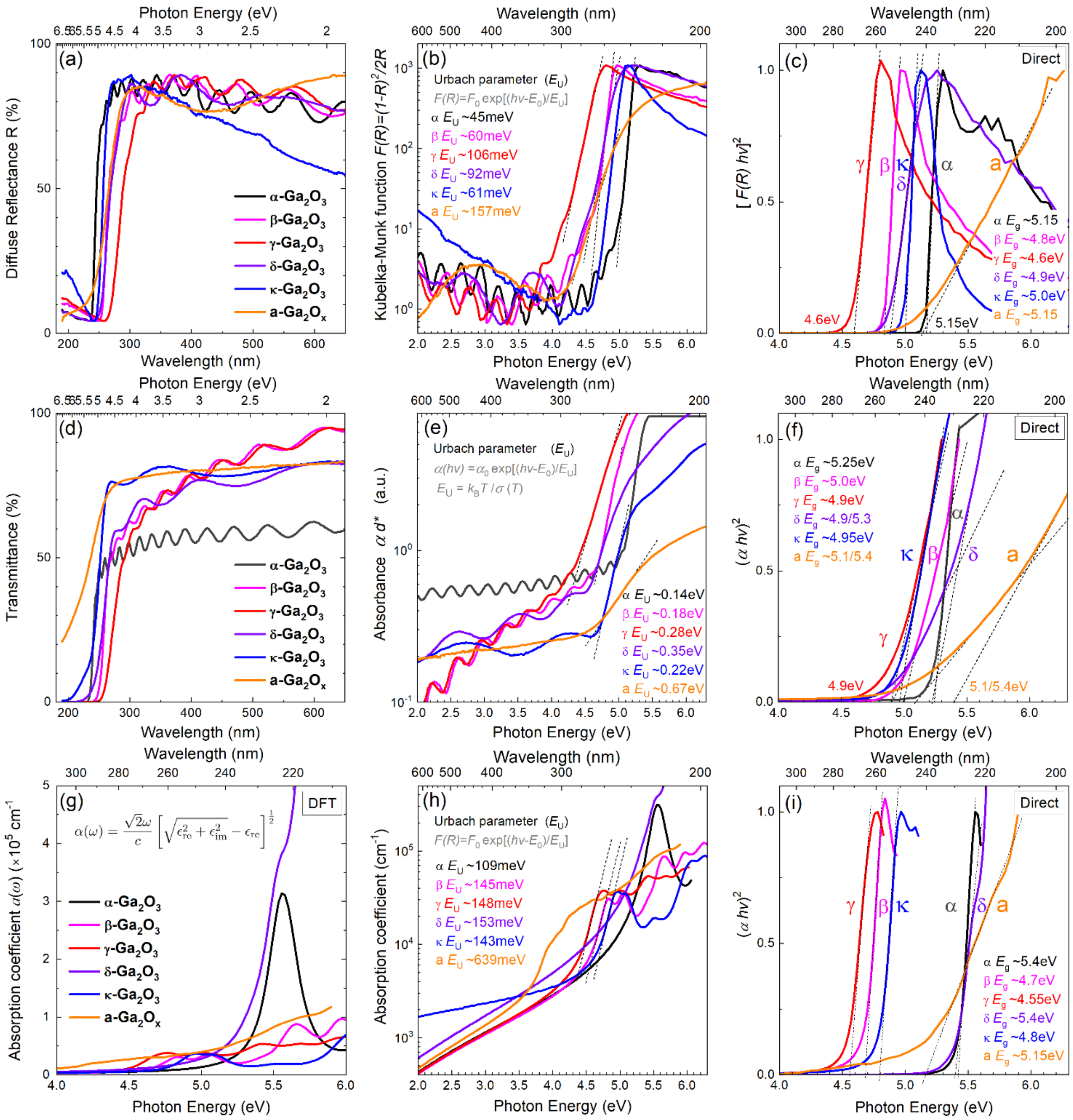

**Figure 3**. Optical bandgap assessment of the $\alpha$-, $\beta$-, $\gamma$-, $\delta$-, and $\kappa$-$Ga_2O_3$ polymorphs, as well as amorphous a-$Ga_2O_x$, from room-temperature diffuse-reflectance and transmittance measurements: (a) Diffuse-reflectance spectra (normalized), (b) Kubelka-Munk function *F(R)* versus incident photon energy *hv*, (c) Tauc plot of the modified Kubelka-Munk function $[F(R)hv]^2$ assuming direct-allowed transitions. (d) Transmittance spectra (normalized), (e) Absorbance ($\alpha$ *d*) as a function of photon energy *hv*. (f) Tauc plot of the absorbance $[\alpha\, hv]^2$ considering direct-allowed transitions. The straight lines are linear extrapolations providing the Urbach energy $E_U$ (from the inverse slopes in panels b, e, h) and the bandgap $E_g$ parameters (from the energy-axis intercepts in panels c, f, i). For comparison with experiment, theoretical absorption spectra obtained from DFT calculations are processed in the same manner and shown in panels (g-i).

Figure 3 summarizes the optical absorption and bandgap results for the $\alpha$, $\beta$, $\gamma$, $\delta$ and $\kappa$ polymorphs of $Ga_2O_3$ , as well as amorphous a-$Ga_2O_x$, derived from diffuse-reflectance and transmittance measurements performed at room temperature. Herein, we follow the common method for estimating bandgaps from the fundamental absorption spectrum, based on the relation $\alpha h\nu \sim (h\nu - Eg)^n$, where $\alpha$ is the absorption coefficient at photon energy $h\nu$ and $n$ depends on the type of optical transition ($n = 1/2$ for direct-allowed and $n = 2$ – for indirect-allowed).[42] Accordingly, direct and indirect bandgaps can be deduced from linear interpolations of $(\alpha h\nu)^2$ and $(\alpha h\nu)^{1/2}$ plotted against photon energy ($h\nu$), which is a graphical analysis approach known as the Tauc method.[43] In the case of transmittance measurements, the spectra are first transformed into absorbance ($A = \alpha d = -\ln T$), which can then be analyzed using Tauc plots, as shown in Figures 3d, 3e, and 3f, respectively. Similarly, the measured diffuse-reflectance values ($R$) are converted to the Kubelka–Munk (K-M) function, $F(R) = (1-R)^2/2R = \alpha/S$, where α is the absorption coefficient, and S is the scattering factor. Assuming $S$ is a constant, K-M function is directly proportional to absorption coefficient, i.e., $F(R) \sim \alpha$, making the K-M spectra analogous to absorbance spectra. Thus, the Tauc method can be applied to diffuse-reflectance measurements by using a modified K-M function, $[F(R)h\nu]^2$, plotted against photon energy ($h\nu$), to assess the bandgap. Figures 3a, 3b, and 3c illustrate the three stages of the diffuse-reflectance data transformation.

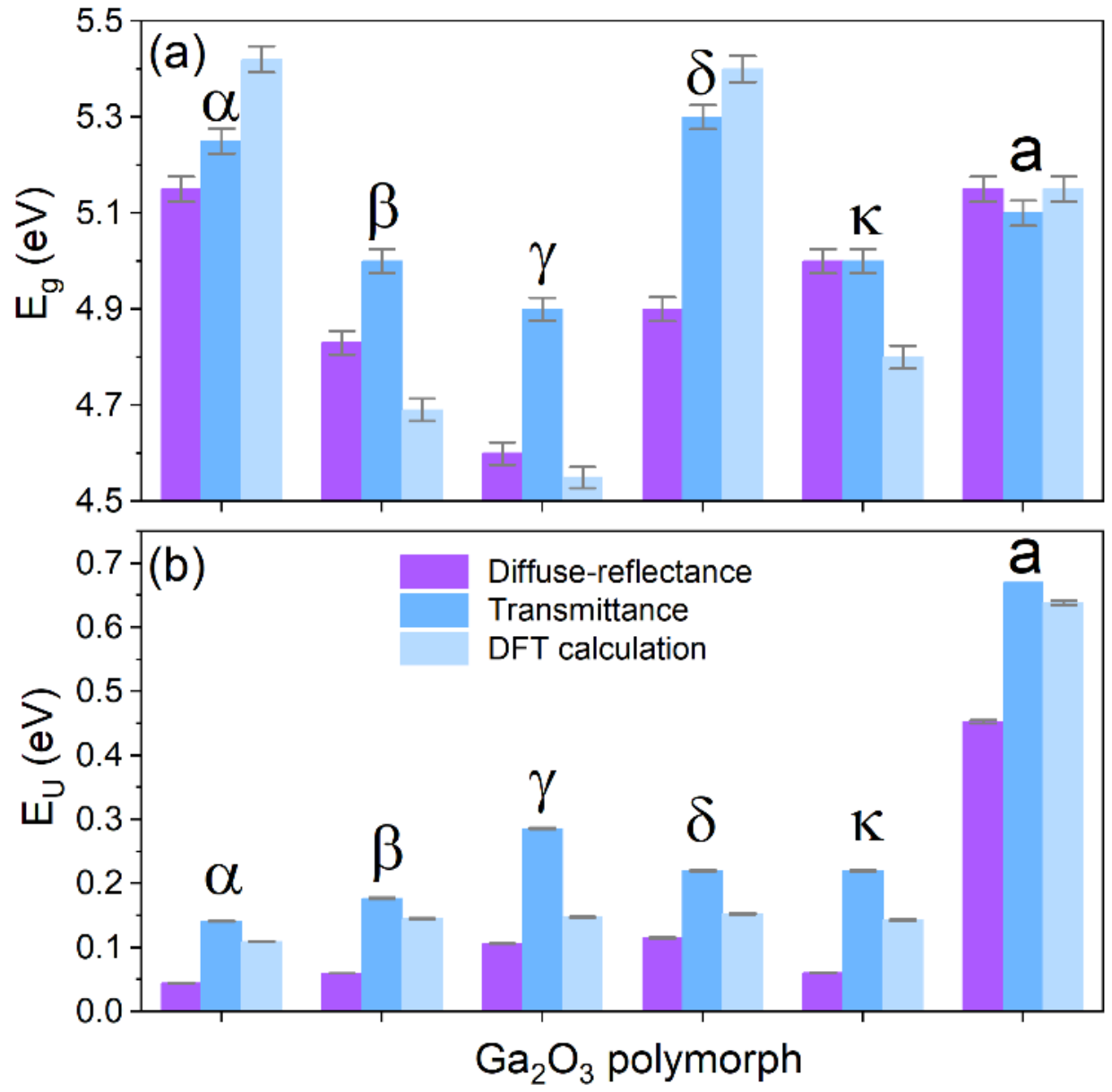

**Figure 4.** Comparison of the (a) optical bandgap and (b) Urbach energy parameters for the $\alpha$-, $\beta$-, $\gamma$-, $\delta$-, and $\kappa$-$Ga_2O_3$ polymorphs, as well as amorphous a-$Ga_2O_x$, obtained from diffuse-reflectance and transmittance measurements, together with values derived from DFT-calculated absorption spectra.

As noted earlier, experimental data for certain $Ga_2O_3$ polymorphs remain limited, while theoretical bandgap predictions are more readily available. To better correlate theory with experiment, we evaluate the optical properties of the $\alpha$-, $\beta$-, $\gamma$-, $\delta$-, and $\kappa$-$Ga_2O_3$ polymorphs, as well as amorphous a-$Ga_2O_x$ using HSE06 hybrid-functional DFT calculations of the frequency-dependent complex dielectric function, as described further in Section 2.5. For direct comparison with experiment, the theoretical absorption spectra are processed using the same analysis procedure as the experimental data and presented in Figs. 3(g-i).

In addition to estimating bandgaps from the fundamental absorption edge, we also examine the region below the energy gap, known as the Urbach tail, which is described by the exponential relation[44] $\alpha(h\nu) = \alpha_0 \exp[(h\nu - E_0)/E_U]$. Here, $E_U$ is the Urbach energy parameter, commonly used to quantify energy disorder at the band edges. This disorder parameter accounts for both structural disorder, which causes localized exponential-tail states, and dynamic disorder from electron-phonon scattering. In Figure 3, the absorption spectra are presented on a semilogarithmic scale; hence, $E_U$ parameters are quantified by linear fitting of the absorption edge slopes indicating the degree of imperfection for each polymorph. The optical bandgap and room-temperature Urbach energy parameters for the $\alpha$, $\beta$, $\gamma$, $\delta$ and $\kappa$-$Ga_2O_3$ polymorphs, as well as amorphous a-$Ga_2O_x$ , are summarized in Table I. The values estimated from the transmittance and diffuse-reflectance measurements are also plotted in Fig. 4 for direct comparison, together with the theoretical predictions from the DFT calculations. One can notice an inverse correlation between the Urbach energy $E_U$ and the optical band gap $E_g$ of the polymorphs. The amorphous a-$Ga_2O_x$ exhibits the widest Urbach tail, as expected for a disordered material. Several other correlations are noteworthy when comparing the two measurement techniques. While polymorph-specific trends are identical, there is an apparent technique-specific discrepancy in under-/over-estimated $E_g$/$E_U$ values. Additionally, there is a structure-specific trend where bulk materials consistently show lower (underestimated) bandgap $E_g$ values compared to thin films (see data in Figures 3 and 4, and Figure S1 in the Supplemental Material). The latter factor most likely is the key contributor to the widely scattered bandgap values in the literature reports. Indeed, as previously mentioned, the most direct way of extracting the optical band gap is to determine the photon energy position at which the linear extrapolation of the absorption edge intersects the baseline. At the band-edge, the absorption coefficient ($\alpha$) reaches the level of $10^5$ to $10^6$ cm$^{-1}$, thus transmission measurements of the entire interband absorption spectrum require the thickness of the media ($d$) to be in the range of 10 to 100 nm, considering optical depth limit $\alpha d \leq 1$. This requirement is rarely met in practice, since optical measurements usually aim to be non-destructive by preserving the original thickness of media, which might differ by several orders of magnitude in the case of bulk crystals (wafers) compared to thin films. Consequently, the apparent onset of the absorption edge varies depending on the actual thickness of the medium. For thick (bulk) material, the bandgap estimates are influenced by the low-energy tail (onset) of the absorbance, leading to underestimated (systematically lower) $E_g$ values compared to those obtained for thin films. The optical bandgap and Urbach energy parameters for primary sample set (thin films) and auxiliary subset (bulk (010) and (-201) $\beta$-$Ga_2O_3$ wafers and $\gamma/\beta$ double polymorph structure) are summarized in Table S1 in the Supplemental Material.

**TABLE I.** Optical bandgap ($E_g$) and Urbach energy ($E_U$) parameters extracted from diffuse-reflectance and transmittance measurements, as well as from density functional theory (DFT) calculations of absorption spectra. The fundamental (thermodynamic) bandgap ($E_g$*), obtained directly from the energy difference between the highest occupied valence-band maximum (VBM) and the lowest unoccupied conduction-band minimum (CBM), is also presented for reference.

| Polymorph | Diffuse Reflectance | | Transmittance | | DFT calculations | | |
|---|---|---|---|---|---|---|---|
| | $E_g$ (eV) | $E_U$ (meV) | $E_g$ (eV) | $E_U$ (meV) | $E_g$ (eV) | $E_U$ (meV) | $E_g$* (eV) |
| α-$Ga_2O_3$ | 5.15 ± 0.03 | 45 ± 0.6 | 5.25 ± 0.03 | 141 ± 0.7 | 5.42 ± 0.02 | 110± 0.5 | 5.4 |
| β-$Ga_2O_3$ | 4.83 ± 0.02 | 60 ± 0.3 | 5.00 ± 0.02 | 177 ± 0.9 | 4.69 ± 0.02 | 145 ± 0.5 | 4.82 |
| γ-$Ga_2O_3$ | 4.60 ± 0.02 | 106 ± 0.5 | 4.90 ± 0.02 | 286 ± 1.4 | 4.55 ± 0.02 | 148 ± 0.5 | |
| δ-$Ga_2O_3$ | 4.75 ± 0.02 | 115 ± 0.5 | 4.90 ± 0.02 | 350 ± 1.4 | 5.4 ± 0.02 | 153 ± 0.5 | 5.11 |
| κ-$Ga_2O_3$ | 5.00 ± 0.02 | 61 ± 0.3 | 5.00 ± 0.02 | 220 ± 1.1 | 4.8 ± 0.02 | 143 ± 0.5 | 4.92 |
| a-$Ga_2O_x$ | 5.15 ± 0.02 | 157/453 ± 1.4/5 | 5.10 ± 0.03 | 670 ± 6.8 | 5.15 ± 0.02 | 639 ± 6.8 | 4.53 |

Finally, it is also noteworthy that the bandgap structure, or rather the selected type of interband optical transitions, can be potential sources of $E_g$ uncertainty. The theoretical studies predict that the fundamental energy gap of $\beta$-$Ga_2O_3$ is indirect, yet due to the minor energy difference between the direct and indirect gaps (~ 30 meV), in practice it is usually regarded as a direct bandgap material.[45,46] Following this convention, the Tauc plots for direct optical transitions are considered for the bandgap analysis in Figure 3. Alternatively, when assuming Tauc plots for indirect transitions, the extracted bandgap values are systematically lower than those for direct transitions, and the discrepancy is considerably greater than that predicted by theory. The reasons are similar to those discussed for bulk versus thin films, as indirect Tauc plots tend to emphasize the early onset of the absorption edge by effectively stretching this region due to the square-root $(\alpha h\nu)^{1/2}$ dependence of the axis.

## 2.3. Emission properties

The emission properties, such as spectral contents and quantum efficiency of photoluminescence (PL), provide an additional set of optical signatures that distinguish different $Ga_2O_3$ polymorphs. Figure 5 displays PL spectra measured at 10K of *α, β, γ, δ, κ* $Ga_2O_3$ polymorphs and amorphous a-$Ga_2O_x$. Here, top panel (Figure 5a) presents normalized PL spectra of the polymorphs for a direct comparison of their emission signatures, whereas the original (raw) spectra are plotted on a semi-logarithmic scale in Figure 5b for relating their quantum efficiencies. The dominant UV emission of the *α, β, γ, δ* and *κ*-$Ga_2O_3$ polymorphs in the range 2.9-3.3 eV share a common intrinsic origin linked to recombination of free electrons and self-trapped holes (STH).[47,48,49,50] By contrast, the characteristic green luminescence (GL) of the amorphous $Ga_2O_x$ is apparently of a different nature, most likely involving a variety of intrinsic defects ($O_i$ , $V_{Ga}$ and $V_{Ga}$-$V_O$) in donor-to-acceptor pair (DAP) recombination processes.[51,52] This association is further supported by comparing the luminescence components and quantum efficiencies of the original (raw) PL spectra shown in Figure 5b. The amorphous a-$Ga_2O_x$ exhibits significantly, by orders of magnitude, lower quantum efficiency compared to crystalline polymorphs and apparently lacks intrinsic (STH) features. Instead, it predominantly exhibits defect-related green-red emission, which is also observed in spectra of all crystalline polymorphs, though not as a dominant characteristic.

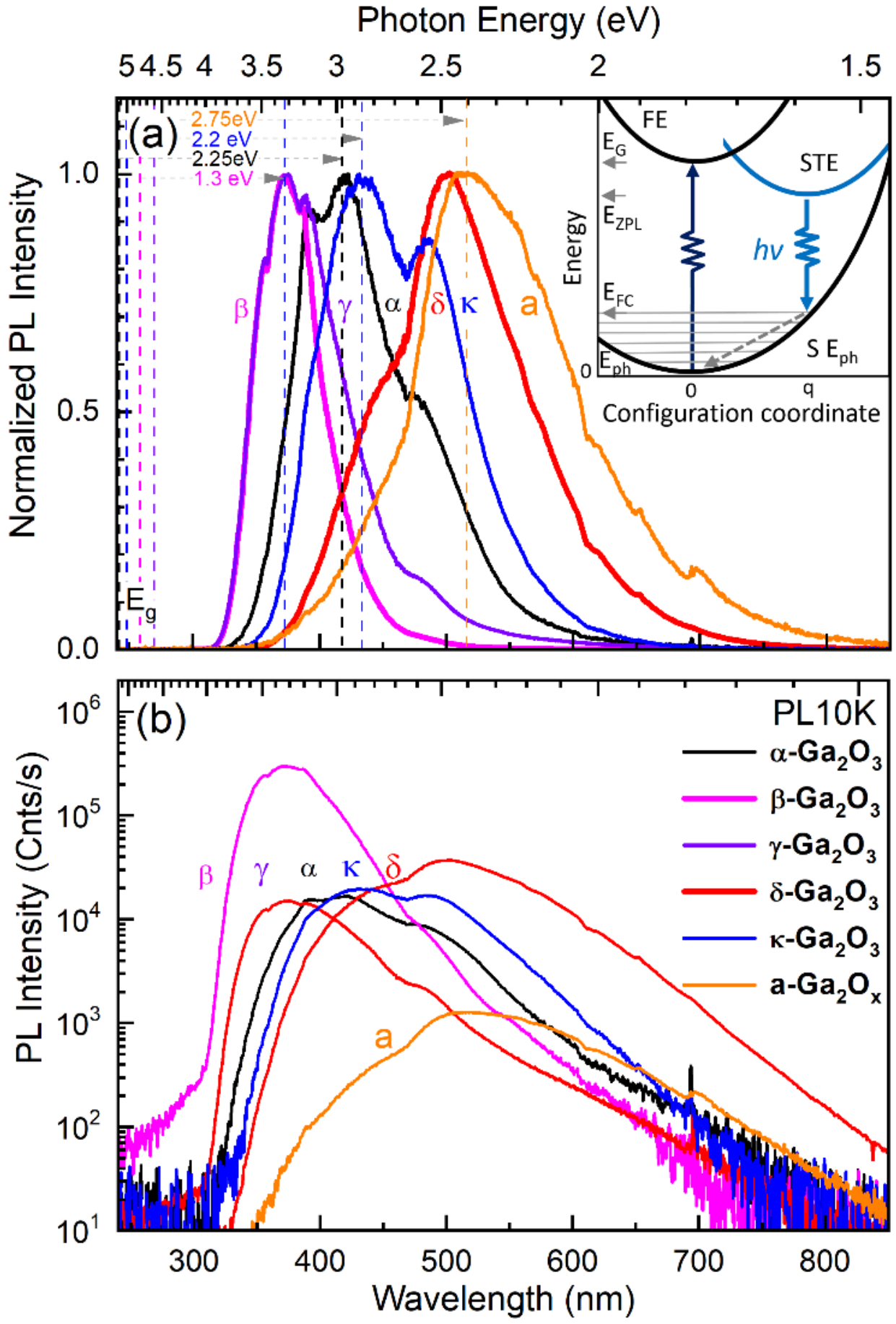

**Figure 5.** Photoluminescence (PL) spectra of the *α*-, *β*-, *γ*-, *δ*-, and *κ*-$Ga_2O_3$ polymorphs, as well as amorphous a-$Ga_2O_x$, measured at 10K. (a) Normalized PL spectra on a linear scale, highlighting characteristic emission features - peak position (PP) and full width at half maximum (FWHM). Vertical dashed lines indicate salient spectral features with indicated energy separation from the band-edge (Stokes shifts). The inset shows a schematic configuration-coordinate (CC) diagram illustrating free exciton (FE) and self-trapped exciton (STE) states (i.e., free electron and self-trapped hole), which underlie the large intrinsic Stokes shifts in $Ga_2O_3$ polymorphs. (b) Raw PL spectra on a semilogarithmic scale, revealing the variations in luminescence efficiency among the polymorphs.

PL results in Figure 5 reveal distinct, clearly identifiable spectral signatures for *α, β, κ* polymorphs, and amorphous a-$Ga_2O_x$, with *β* and *γ* phases exhibiting more closely matching signatures. The energy difference between the bandgap ($E_g$) and the emission peak position (PP), known as the Stokes shift ($E_S$), is indicated by arrows in Figure 5a. The cause of the shift for intrinsic (STH) emission is illustrated in the configuration-coordinate (CC) diagram in the Inset of Figure 5b, with further details on CC models provided in Refs. [53,54]. One can observe that Stokes shift $E_S$ scales with the bandgap $E_g$ of a particular polymorph, i.e., *β* phase with the narrowest $E_g$ (4.65 eV) also exhibits minimal $E_S$ (1.37 eV), whereas *α* phase with the broadest $E_g$ (5.15 eV) shows the largest $E_S$ (2.25 eV).

For all polymorphs, the large Stokes shift ($E_S > 1$ eV) and broad STH emission width ($W_{FWHM} > 0.5$ eV) indicate strong electron-phonon (*e-ph*) coupling. This is characterized by a Huang-Rhys factor[55] exceeding $S$ >20, as follows from the relationships[54] $W_{FWHM} \sim 2.35\, E_{ph}\, S^{1/2}$ and $E_S \sim E_{ph}\, S$ assuming the average phonon energy $E_{ph}$ ~ 45 meV.[50,56] A more accurate evaluation of *e-ph* coupling parameters can be achieved by fitting the STH emission as outlined below. In addition to general PL signatures of the polymorphs, such as dominant peak position and width, further analysis includes Gaussian deconvolution of emission spectra to identify overlapping components. The singled out components can then be analyzed and compared to known optical signatures from the literature. Figure 6 presents PL spectra of different polymorphs with deconvolution components color-coded and labeled in accord to common notation of $Ga_2O_3$ luminescence bands in literature: ultraviolet (UVL) (3.2 - 3.6 eV),[48,49,50,57] blue (BL) (2.8 - 3.0 eV),[58,59] green (GL) (2.4 - 2.5 eV)[60] and red (RL) (1.7 - 1.9 eV).[61,62,63] The fitting and emission band parameters are listed in Table II. The candidate transitions predicted by theoretical calculations[50] are indicated by vertical markers in Figure 6d. In addition, the bold black curves in Figures 6b, 6d, and 6f show the fitting of the intrinsic STH emission component using the CC-model in the framework proposed in Ref. [64], with the parameters detailed in Table III. It is important to note that a single CC-model fit can replace several (up to three) Gaussian deconvolution components in the blue-UV region of the spectrum.

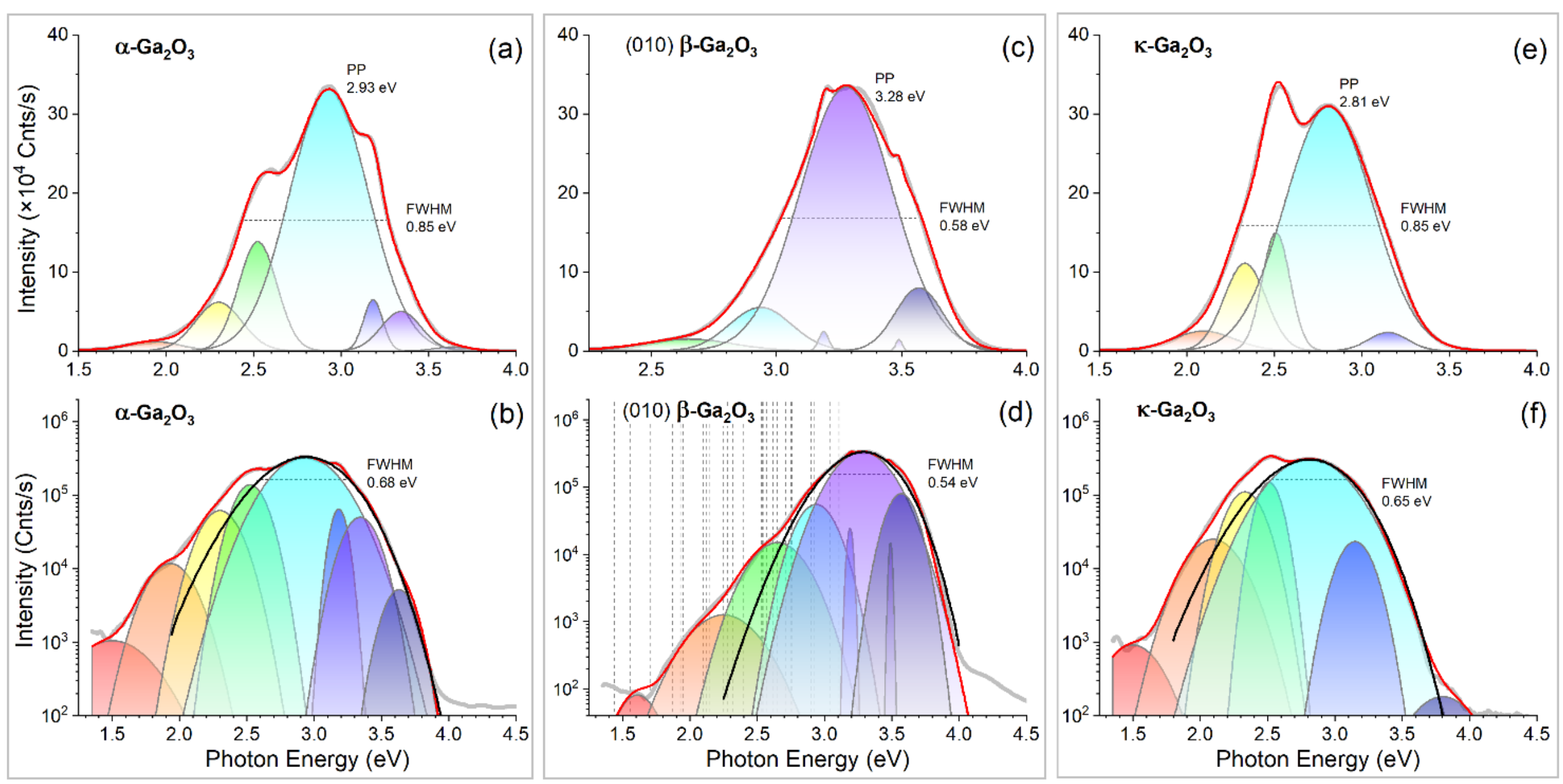

**Figure 6.** Deconvolution of the PL spectra for rhombohedral $\alpha$-$Ga_2O_3$ , monoclinic $\beta$-$Ga_2O_3$ and orthorhombic $\kappa$ -$Ga_2O_3$ polymorphs. Representative spectra are shown on linear and semilogarithmic scales in panels (a,c,e) and (b,d,f), respectively. Gaussian deconvolution components are represented by grey curves and shaded areas, the corresponding fitting parameters are summarized in Table II. Vertical dashed lines in panel (d) mark optical transitions that have been experimentally established or theoretically predicted in the literature reports.[57-63]

**TABLE II.** Gaussian deconvolution and fitting parameters of PL emission bands of $\alpha$- , $\beta$- and $\kappa$-$Ga_2O_3$ polymorphs. The main emission components, numbered from 1 to 8, are labeled following common assignments in the literature for red (RL), green (GL), blue (BL) and ultraviolet (UVL) luminescence bands.[57-63] *PP* stands for peak position, $W_{FWHM}$ - full width at half maximum, *I*(norm) - normalized intensity of emission components.

| Gaussian component / Band assignment | | $\alpha$ -$Ga_2O_3$ | | | $\beta$ -$Ga_2O_3$ | | | $\kappa$-$Ga_2O_3$ | | |
|---|---|---|---|---|---|---|---|---|---|---|
| | | *PP* (eV) | $W_{FWHM}$ (eV) | *I*(norm) | *PP* (eV) | $W_{FWHM}$ (eV) | *I*(norm) | *PP* (eV) | $W_{FWHM}$ (eV) | *I*(norm) |
| 1 | RL | 1.5 | 0.5875 | 0.00349 | 1.61 | 0.29375 | 1.6129E-4 | 1.5 | 0.41125 | 0.00206 |
| 2 | RL | 1.93 | 0.3525 | 0.02362 | 2.25 | 0.50525 | 0.00435 | 2.1 | 0.41125 | 0.05672 |
| 3 | GL | 2.3 | 0.31725 | 0.11014 | 2.65 | 0.41125 | 0.04225 | 2.33 | 0.282 | 0.17211 |
| 4 | BL | 2.52 | 0.2585 | 0.20409 | 2.94 | 0.29375 | 0.11157 | 2.51 | 0.188 | 0.15469 |
| 5 | BL | 2.93 | 0.52875 | 1 | 3.19 | 0.047 | 0.00792 | 2.81 | 0.5875 | 1 |
| 6 | BL | 3.18 | 0.12925 | 0.04778 | 3.28 | 0.43475 | 1 | 3.15 | 0.27025 | 0.03431 |
| 7 | UVL | 3.34 | 0.27025 | 0.07686 | 3.49 | 0.02938 | 0.00296 | 3.65 | 0.235 | 0 |
| 8 | UVL | 3.63 | 0.235 | 0.00696 | 3.57 | 0.22325 | 0.12265 | 3.8 | 0.47 | 4.64E-4 |

**TABLE III.** Effective fitting parameters of STH emission bands of $\alpha$- , $\beta$- and $\kappa$-$Ga_2O_3$ polymorphs assuming one-dimensional configuration-coordinate model:[64] *S* is Huang-Rhys factor, $E_0$ is zero phonon line energy, $E_{ph}$ is phonon energy, *PP* is peak position and $W_{FWHM}$ – width (full width at half maximum) of the STH emission band.

| Polymorph | *PP* (eV) | $W_{FWHM}$ (eV) | *S* | $E_0$ (eV) | $E_{ph}$ (meV) |
|---|---|---|---|---|---|
| $\alpha$ -$Ga_2O_3$ | 2.93 | 0.68 | 38.5 | 4.65 | 45 |
| $\beta$ -$Ga_2O_3$ | 3.28 | 0.54 | 24.5 | 4.37 | 45 |
| $\kappa$ -$Ga_2O_3$ | 2.81 | 0.65 | 38 | 4.5 | 45 |

### 2.4. Near-field scanning nano-FTIR analysis

The near-field optical signatures of the polymorphs were identified using scanning nanoscale Fourier Transform Infrared (nano-FTIR) spectroscopy. This involved scanning the cross-section of the double $\gamma/\beta$ polymorph structure (Figure 7) and analyzing the plan-view surfaces of the single-phase polymorph arrangements at the nanoscale (Figure 8). The nano-FTIR technique combines scattering-type scanning near-field optical microscopy (s-SNOM) with broadband IR laser illumination and the detection of elastically scattered IR light using an interferometric Fourier transform-based scheme. [65,66] The focusing of the broadband IR laser beam on a metallized AFM tip and the ensuing confinement of light at its apex facilitate localized near-field probing with a resolution of 10 nm. Thus, nano-FTIR enables true FTIR spectroscopy at the spatial resolution of AFM, allowing for nanoscale chemical identification and hyperspectral imaging. Referring to conventional (far-field) FTIR studies of $\beta$-$Ga_2O_3$, a number of bands observed in the 620 - 725 $cm^{-1}$ region are typically assigned to Ga-O stretching vibrations.[57,67,68,69] Generally, a direct comparison of near-field and far-field FTIR fingerprints is not trivial, though it has proven effective for nonmetal materials like polymers.[65] Nonetheless, certain insights can be gained by comparing our nano-FTIR results with the signatures of $\beta$- and $\kappa$-$Ga_2O_3$ polymorphs reported in Ref. [70]. The far-field FTIR signature of $\beta$-$Ga_2O_3$ is linked to optical phonon modes at 673, 692 and 732 $cm^{-1}$, which compare well with the nano-FTIR features at 666, 670, 696 and 728 $cm^{-1}$ in Figure 8b. The FTIR signature of $\kappa$-$Ga_2O_3$ is associated with a broadened peak due to several merged optical phonon modes at 715 $cm^{-1}$, whereas nano-FTIR demonstrates a distinct mode at 725 $cm^{-1}$ in Figure 8d. The near-field optical signatures in Figures 7 and 8 align with existing literature on far-field FTIR features for $\beta$ and $\kappa$ polymorphs, while also extending observations to a previously unreported for $Ga_2O_3$ nanometer-scale spatial domain. Moreover, the nano-FTIR signatures collected for all polymorphs serve as a reference point for future research, making them a valuable addition to the optical library presented in this study.

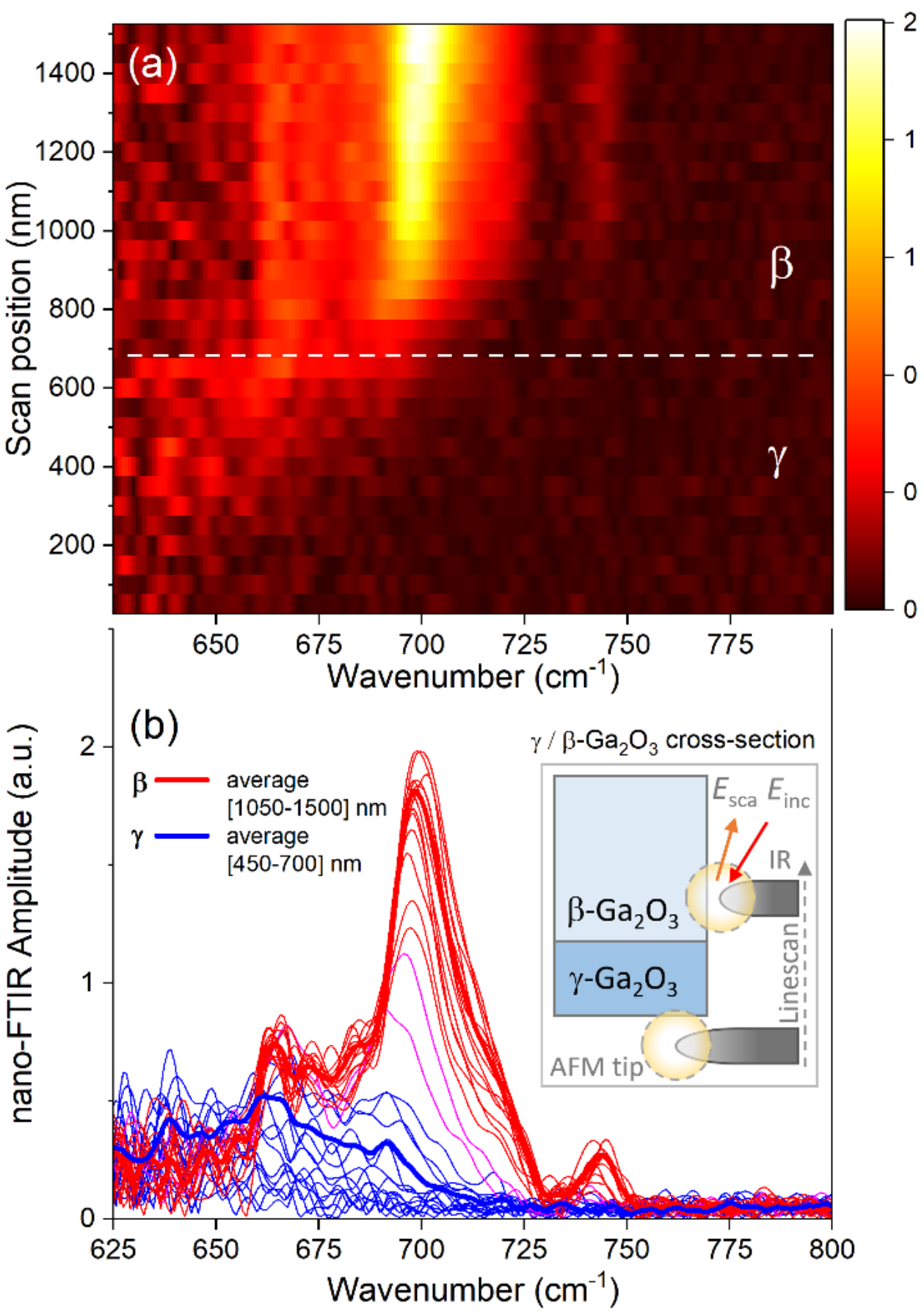

**Figure 7.** Nanoscale structural and spectroscopic characterization of the double-polymorph $\gamma/\beta$-$Ga_2O_3$ structure using nano-FTIR. (a) Two-dimensional (2D) representation of a cross-sectional nano-FTIR line scan acquired over 1500 nm with a 50 nm step size. (b) Near-field optical signatures of the $\gamma$- and $\beta$-$Ga_2O_3$ polymorphs represented by averaged spectra in the respective domains and shown as bold lines.

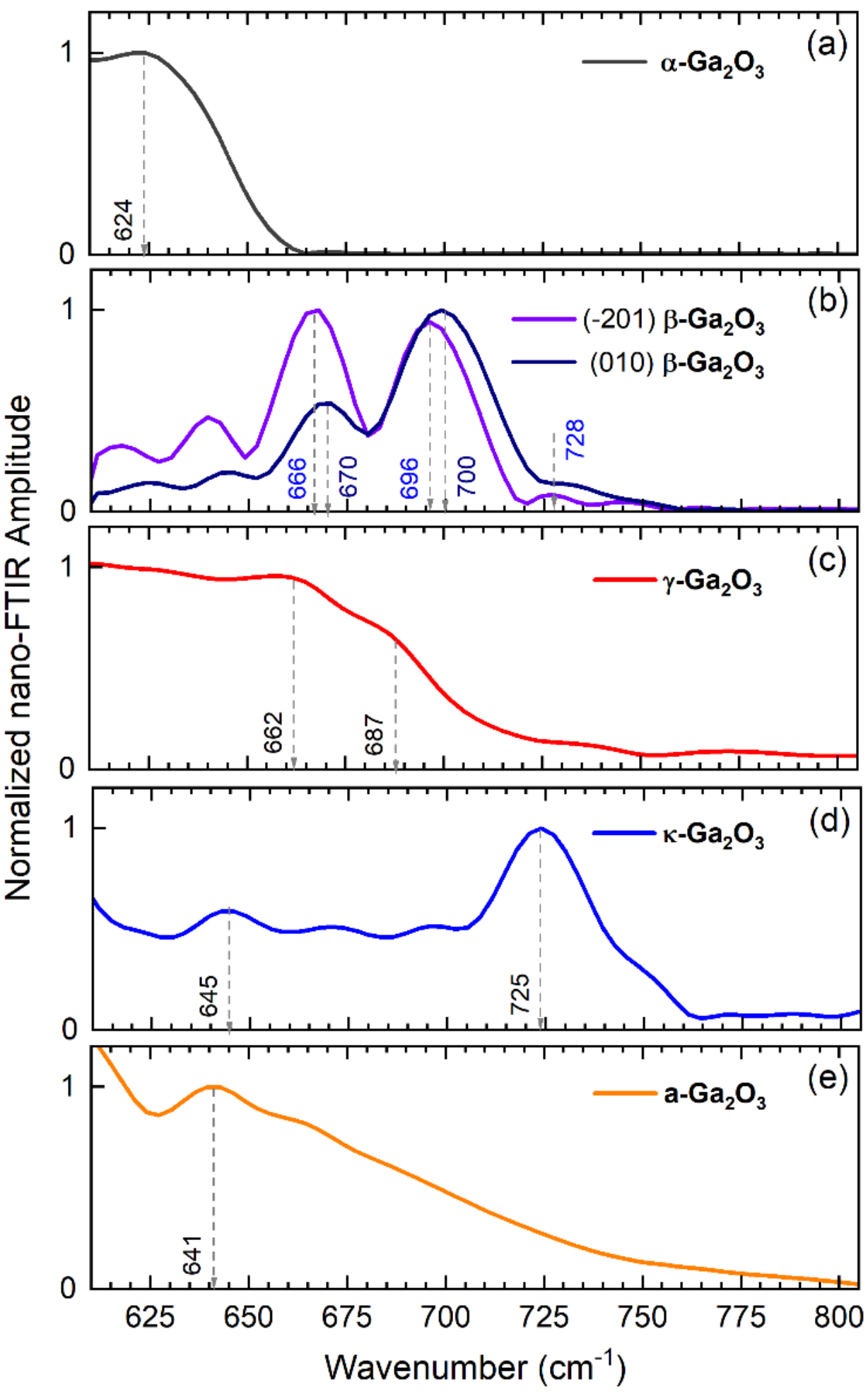

**Figure 8.** Near-field optical signatures obtained by nano-FTIR for the $\alpha$-, $\beta$-, $\gamma$-, and $\kappa$-$Ga_2O_3$ polymorphs , as well as amorphous a-$Ga_2O_x$. Vertical dashed lines indicate the salient spectral features characteristic of each polymorph.

### 2.5. DFT calculations

The ground-state lattice parameters of the five polymorphs and amorphous cells are listed in Table IV. It is well-known that HSE06 hybrid functional can predict the smaller lattice constants which is in a better agreement with the experimental values, comparing to generalized gradient approximation *xc* functional.[84,85] Figure 9 summarizes the key optical properties of the $Ga_2O_3$ polymorphs obtained from DFT calculations of the frequency-dependent complex dielectric function.

**TABLE IV.** Optimized lattice constants (Å) and inter-axial angles (°) obtained from HSE06 hybrid-functional DFT calculations. For *γ*-$Ga_2O_3$ and amorphous a-$Ga_2O_x$, the exact lattice constants are not well-defined due to inherent structural disorder; therefore, an averaged value over three random cells is shown here. For the amorphous $Ga_2O_3$ cell, the atomic ratio is kept exactly at Ga : O = 2 : 3.

| Polymorph | a | b | c | ∠α | ∠β | ∠γ |
|---|---|---|---|---|---|---|
| *α* -$Ga_2O_3$ 30 atoms hexagonal | 4.970 | | 13.412 | | 90 | 120 |
| *β* -$Ga_2O_3$ 20 atoms monoclinic | 12.231 | 3.035 | 5.797 | 90 | 103.848 | 90 |
| *γ* -$Ga_2O_3$ 160 atoms 1 × 1 × 3 cubic | | 8.219 | | | 90 | |
| *δ* -$Ga_2O_3$ 80 atoms cubic | | 9.401 | | | 90 | |
| *κ* -$Ga_2O_3$ 40 atoms orthorhombic | 8.658 | 5.026 | 9.265 | | 90 | |
| a-$Ga_2O_3$ 80 atoms cubic | | 10.008 | | | 90 | |

The fundamental (thermodynamic) bandgaps of 5.40 eV, 4.82 eV, 5.11, 4.92 eV, and 4.53 eV are theoretically predicted for the *α*-, *β*-, *δ*-, *κ*-, and amorphous (a-$Ga_2O_x$) phases, respectively. The bandgap of the *γ* phase is not universally defined owing to the inherent disordering. For the three amorphous cells used in calculations, with an average lattice constant of 10.008 Å, their fundamental (thermodynamic) bandgaps are 4.53541 eV, 4.56426 eV, and 4.50106 eV, respectively (average 4.53 eV). For the amorphous cells, there are actually one or two additional dependent parameters. The first is the cell density: for a relatively high-density cell (obtained by compressing the amorphous cell to a new local minimum via elastic deformation), the bandgap increases by about 0.3 eV. The second concerns defect levels. In an amorphous cell, the most likely defect level (where the forbidden band is not perfectly clean) arises from undercoordinated Ga. As a result, the hybrid defect band (mainly contributed by Ga-4s orbitals from undercoordinated Ga) lies well below the continuous conduction band by approximately 0.5 eV. The values reported above exclude the effects of such defect levels.

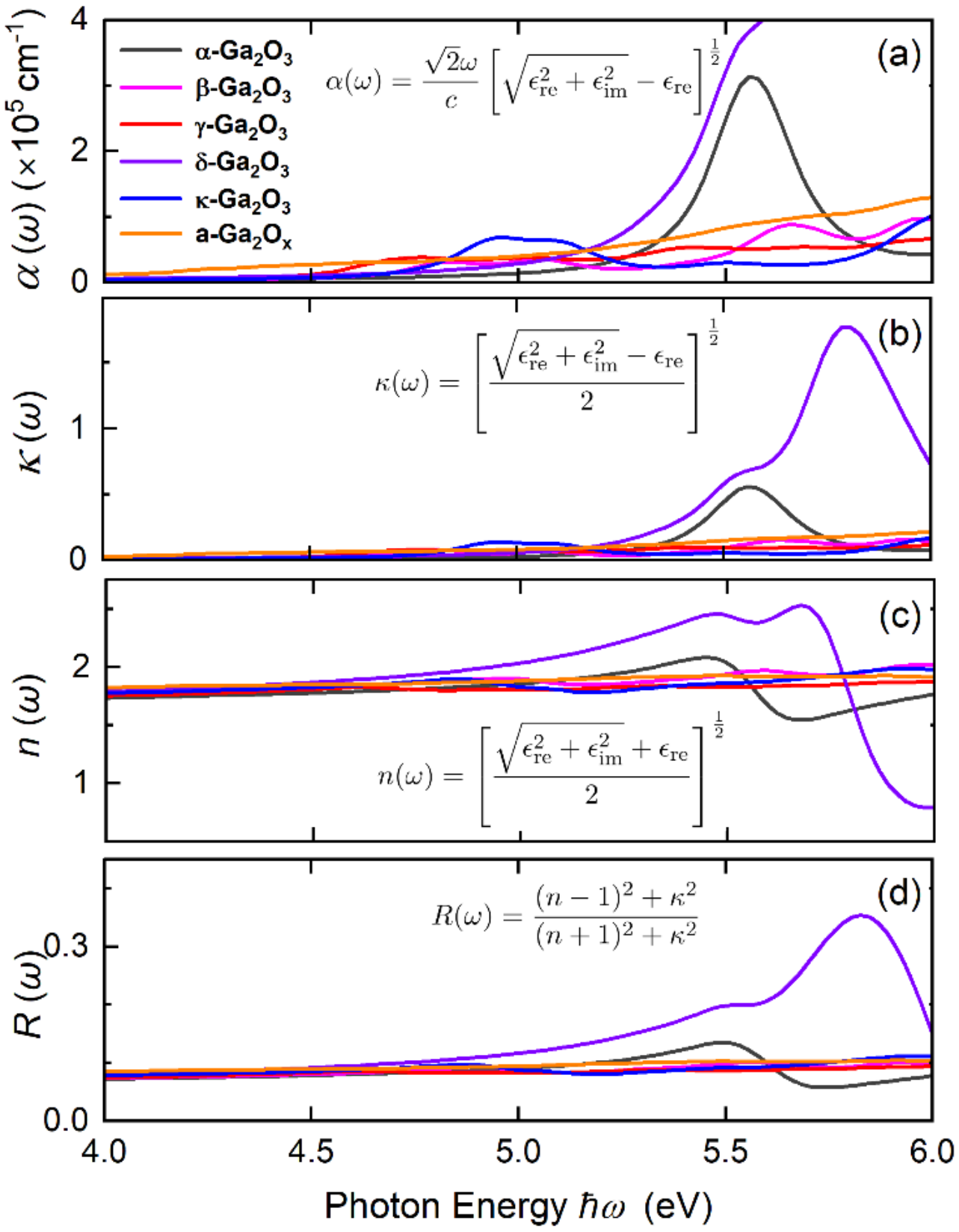

**Figure 9.** Optical properties of the *α*-, *β*-, *γ*-, and *κ*-$Ga_2O_3$ polymorphs, as well as amorphous a-$Ga_2O_x$, obtained from HSE06 hybrid-functional DFT calculations of the frequency-dependent complex dielectric function. Shown are the calculated (a) absorption coefficient, (b) extinction coefficient, (c) refractive index, and (d) reflectivity.

In the DFT framework, the fundamental (thermodynamic) bandgaps are obtained directly from the energy difference between the highest occupied band (VBM) and the lowest unoccupied band (CBM). This provides a unique opportunity to assess how these bandgaps compare with those derived from DFT-calculated absorption spectra using the Tauc method, a conventional approach in optical characterization. Both sets of theoretical bandgap estimates for the polymorphs are summarized in Table I, and also shown alongside the corresponding experimental values in Figs. 3 and 4. Overall, the theoretical predictions align well with the experimental data.

## 3. Conclusion

In this work, we collected data that enables a reliable comparison of the optical signatures of different $Ga_2O_3$ polymorphs. To achieve this, we cross-correlated optical emission and absorption signatures in a systematic set of thin-film samples of the *α*-, *β*-, *γ*-, *δ*-, and *κ*-phases, complemented by differently oriented *β*-phase bulk crystals and *γ/β* double-polymorph structures. We demonstrate that the optical bandgap and emission features scale consistently across these $Ga_2O_3$ polymorphs when methodological uncertainties are minimized through systematic sample selection and unified analysis procedures. This work highlights and quantifies systematic differences arising from the choice of methods and techniques employed in material characterization, including theoretical calculations that are processed in the same manner as the experimental data. Furthermore, to extend conventional far-field optical phase identification down to the nanoscale, we obtained near-field optical signatures of $Ga_2O_3$ polymorphs via nano-FTIR reported herein for the first time. These measurements provide a nanoscale reference for future studies and constitute a valuable addition to the optical library presented in this study. Overall, the present dataset comprise a comprehensive collection of near- and far-field optical polymorph signatures, intended for use by the multidisciplinary research community working with $Ga_2O_3$.

## 4. Experimental Section

### 4.1. Samples

The representative samples of rhombohedral ($\alpha$), monoclinic ($\beta$), cubic defect spinel ($\gamma$), cubic bixbyite ($\delta$) and orthorhombic ($\kappa$) $Ga_2O_3$ polymorphs studied in this work comprise thin films grown on sapphire, disorder-induced ordered layers and bulk single crystals acquired from different sources as described in what follows. The samples were categorized into two sets: primary and auxiliary. The primary set comprises polymorphs formed as thin films on transparent substrates. This configuration enables both transmittance and diffuse-reflectance analysis and the opportunity to directly compare the results:

$\alpha$-$Ga_2O_3$ hetero-epitaxial layers (~1 µm-thick) grown by halide vapor phase epitaxy (HVPE) on c-axis sapphire ((0001) $Al_2O_3$) substrates. The details of synthesis are presented elsewhere [71].

$\beta$-$Ga_2O_3$ hetero-epitaxial layers (~1µm-thick) grown by halide vapor phase epitaxy (HVPE) on c-axis sapphire substrates.[71]

$\gamma$-$Ga_2O_3$ layer (~1 µm-thick) obtained by disorder-induced ordering of β-$Ga_2O_3$ film grown on sapphire. The irradiation with 1.5 MeV $^{58}$Ni+ ions to a dose of $1\times10^{16}$ cm$^{-2}$ was performed at room temperature.

$\delta$-$Ga_2O_3$ hetero-epitaxial ~300 nm-thick layer ($\delta$+$\kappa$ mixed-phase) grown by MOCVD on c-axis sapphire ((0001) $Al_2O_3$) substrate. [86]

$\kappa$-$Ga_2O_3$ hetero-epitaxial layers (150 nm-thick) grown by HVPE on c-axis sapphire ((0001) $Al_2O_3$) substrates.

a-$Ga_2O_X$ amorphous (~1 µm-thick) layers deposited on c-axis sapphire by room-temperature RF magnetron sputtering.

The second set of samples is auxiliary and consists of bulk materials, including single crystals with various orientations and phase-transformed layers within the bulk crystals:

$\beta$-$Ga_2O_3$ (010), (001) and (-201) oriented β-$Ga_2O_3$ single crystal commercial wafers (Tamura Corp., Japan).

$\gamma$-$Ga_2O_3$ layers (300 nm-thick) obtained by disorder-induced ordering of (010) and (−201) β-$Ga_2O_3$ single crystal wafers (Tamura Corp., Japan) resulting in double polymorph $\gamma/\beta$ $Ga_2O_3$ heterostructures.[14]

### 4.2. Methods

*X-ray diffraction*

X-ray diffraction (XRD) 2theta measurements were performed using the RIGAKU SmartLab diffractometer with high-resolution CuKa1 radiation and Ge(440) four-bounced monochromator.

*TEM analysis*

Transmission Electron Microscopy (TEM) investigations were conducted on an FEI Titan G2 60-300 kV equipped with a CEOS DCOR probe-corrector and monochromator. Observations were performed at 200 kV and electron transparent TEM samples with a cross-sectional wedge geometry were prepared by mechanical grinding and polishing (Allied MultiPrep). Final thinning was performed by Ar ion milling with a Fishione Model 1010, and plasma cleaning was applied directly before the TEM investigations, with a Fishione Model 1020.

*Transmittance and Diffuse-Reflectance Spectroscopy*

Optical transmittance and diffuse-reflectance spectroscopy (DRS) measurements were performed at room temperature by employing UV−Vis-NIR spectrophotometer EVO-600 (Thermo Fisher Scientific, Inc.) equipped with Praying Mantis™ diffuse-reflection accessory (DRA), which incorporates two 90° off-axis ellipsoid mirrors that form a highly efficient illumination and collection system. DRA provides results that are qualitatively similar to more common diffuse-reflectance accessory - an integrating sphere - but with the advantages horizontal mounting of samples, and collimation of probing beam to ~2 mm spot without loss of performance. Of particular note is the micrometer-style height adjustment, which enables fine tuning of the focal-plane position to maximize signal from a designated subsurface region - an essential feature for studies of ion-implanted layers and stacked thin films.

The optical bandgaps of the polymorphs were estimated from the onset of the band-edge absorption by employing standard Kubelka-Munk and Tauc methods.[43,72,73]

*Photoluminescence Spectroscopy*

PL spectroscopy measurements were performed at 10K temperature by employing a closed-cycle He refrigerator system (CCS-450 Janis Research, Inc.). The photo-excitation at 246 nm wavelength (5.04 eV) and 10 mW average power was provided by third-harmonic of a pulsed Ti:sapphire laser operating at 80 MHz in a femtosecond mode-locked mode (Spectra-Physics, Tsunami HP and GWU-UHG-23). PL emission was collected by a microscope and analyzed by a fiber-optic spectrometer (Avantes, AvaSpec-Mini3648-UVI25) in the wavelength range 200-1100 nm with spectral resolution below 2 nm.

*Scanning nano-FTIR Spectroscopy*

The near-field optical signatures of the polymorphs and their structural arrangements on a nanoscale were analyzed by scanning Fourier transform infrared nano-spectroscopy (nano-FTIR) technique. Near-field nano-FTIR measurements were carried out at room temperature using a commercial IR-neaSCOPE$^{+s}$ system (attocube systems AG) equipped with a broadband mid-infrared laser source and providing chemical analysis and field mapping at 10 nm spatial resolution. The nanoscale tip-enhanced IR absorption (reflection) spectra, represented by the second harmonic of the imaginary part (amplitude) of the complex spectrum, were acquired in the spectral range 620–1400 cm$^{-1}$ with resolution of 3 cm$^{-1}$ and 16 cm$^{-1}$.

*Computational Details*

All *ab initio* calculations were conducted based on density functional theory (DFT) implemented with the Vienna *Ab initio* Simulation Package (VASP) [74]. A high plane-wave cutoff energy of 700 eV was used to obtain well-converged results. The energy convergence criterion was $1\times10^{-6}$ eV for the electronic self-consistent field iteration, and the force convergence criterion was $5\times10^{-3}$ eV/Å for ionic structural optimization. The first Brillouin zone was sampled with the dense Γ-centered reciprocal Monkhorst−Pack grid [75] with *k*-spacing of 0.15 Å$^{-1}$. Gaussian approximated smearing with a width of 0.03 eV was used to determine the partial occupancies. The Heyd–Scuseria–Ernzerhof (HSE06) hybrid functional [76] with 32% of the exact Hartree–Fock exchange [77] was used throughout structure optimization and calculating electronic and optical properties. It has been tested that this mixing parameter results in bandgaps of 5.40 eV, 4.82 eV, 5.11, 4.92 eV, and 4.53 eV for $\alpha$, $\beta$, $\delta$, $\kappa$, and amorphous a-$Ga_2O_3$ phases, respectively. The bandgap of the $\gamma$ phase is not universally defined owing to the inherent disordering. The mentioned thermodynamic bandgaps are directly extracted from the energy difference between the highest occupied band (VBM) and the lowest unoccupied band (CBM); this explains why they may differ from those expected from optical methods.

Some of the post-processing tasks were done with VASPKIT package.[78] OVITO [79] and VESTA [80] packages were used for visualizing the atomic configurations and charge densities. After the electronic ground states of the optimized cells have been determined, the frequency dependent dielectric matrix is extracted with a Lorentzian broadening of 0.1 eV.[81] The large number of empty conduction band states is tested to reach a converged result. The number of grid points for the electronic density of states (DOS) and dielectric function is set to 2000.

The frequency dependent complex dielectric matrix is defined as [82]:

$$\varepsilon(\omega) = \varepsilon^{re}(\omega) + i\varepsilon^{im}(\omega), \tag{1}$$

where $\varepsilon^{re}(\omega)$ and $\varepsilon^{im}(\omega)$ are the real and imaginary parts, respectively; $\omega$ is the photon frequency in the dimension of an energy as a common notation. The 3 × 3 Cartesian tensor $\varepsilon_{ab}^{im}(\omega)$ of the imaginary part $\varepsilon^{im}(\omega)$ is determined by a summation over empty states using the equation [81]:

$$\varepsilon_{ab}^{im}(\omega) = \frac{4\pi^2 e^2}{\Omega}\lim_{q\to 0}\frac{1}{q^2}\sum_{c,v,\mathbf{k}} 2w_{\mathbf{k}}\delta(\boldsymbol{\epsilon}_{c\mathbf{k}} - \boldsymbol{\epsilon}_{v\mathbf{k}} - \omega) \times \tag{2}$$

$$\times \langle u_{c\mathbf{k}+\boldsymbol{e}_a q}|u_{v\mathbf{k}}\rangle\langle u_{c\mathbf{k}+\boldsymbol{e}_b q}|u_{v\mathbf{k}}\rangle^*,$$

where $\Omega$ is the volume of the primitive cell; $q$ is the Bloch vector norm of the incident wave; the indices $c$ and $v$ are restricted to the conduction and the valence band states, respectively; $\boldsymbol{\epsilon}_{c\mathbf{k}}$ and $\boldsymbol{\epsilon}_{v\mathbf{k}}$ are the energy levels of the corresponding band states with the wave factor $\mathbf{k}$; the vectors $\boldsymbol{e}_a$ and $\boldsymbol{e}_b$ (and omitted $\boldsymbol{e}_c$) are unit vectors for the three Cartesian directions; and finally, $u_{c\mathbf{k}}$ and $u_{v\mathbf{k}}$ are the cell periodic parts of the corresponding band states at the wave factor $\mathbf{k}$. It is also possible to restrict the wave factor $\mathbf{k}$ to the irreducible wedge ($w_{\mathbf{k}}$) of the first Brillouin zone of the crystal symmetry group.

The real part of the dielectric tensor, $\varepsilon^{re}(\omega)$, is obtained by the usual Kramers-Kronig transformation:

$$\varepsilon_{ab}^{\mathrm{re}}(\omega) = 1 + \frac{2}{\pi}P\int_0^\infty \frac{\varepsilon_{ab}^{\mathrm{im}}(\omega')\omega'}{\omega'^2 - \omega^2 + i\eta}d\omega', \tag{3}$$

where $P$ denotes the principle value, and the complex shift $\eta$ is used for spectral broadening. The frequency-dependent linear optical spectra, such as orientation-dependent absorption coefficient, $\alpha_{ab}(\omega)$, can be calculated from the dielectric matrix [82]:

$$\alpha_{ab}(\omega) = \frac{\sqrt{2}\omega}{c}\left[\sqrt{{\varepsilon_{ab}^{\mathrm{re}}}^2 + {\varepsilon_{ab}^{\mathrm{im}}}^2} - \varepsilon_{ab}^{\mathrm{re}}\right]^{1/2}, \tag{4}$$

and reflectivity, $R_{ab}(\omega)$, can be obtained with the formula:

$$R_{ab}(\omega) = \frac{[n_{ab}(\omega) - 1]^2 + \kappa_{ab}(\omega)^2}{[n_{ab}(\omega) + 1]^2 + \kappa_{ab}(\omega)^2}, \tag{5}$$

where $n_{ab}(\omega)$ and $\kappa_{ab}(\omega)$ are the refractive index and extinction coefficient, respectively, with the formulas:

$$n_{ab}(\omega) = \left[\frac{\sqrt{{\varepsilon_{ab}^{\mathrm{re}}}^2 + {\varepsilon_{ab}^{im}}^2} + \varepsilon_{ab}^{\mathrm{re}}}{2}\right]^{1/2} \quad \kappa_{ab}(\omega) = \left[\frac{\sqrt{{\varepsilon_{ab}^{re}}^2 + {\varepsilon_{ab}^{im}}^2} - \varepsilon_{ab}^{re}}{2}\right]^{1/2} \tag{6}$$

## Supporting Information

Supporting information is available from the Online Library or from the author.

## Acknowledgements

The Research Council of Norway is acknowledged for the support to the Norwegian Micro- and Nano-Fabrication Facility, NorFab (project No. 295864), the Norwegian Centre for Transmission Electron Microscopy, NORTEM (project No. 197405), GOFIB (project No. 337627), and DIOGO (project No. 351033). The international collaboration was in part enabled by the INTPART Program funded by the Research Council of Norway (project No. 322382) and UTFORSK Program at the Norwegian Directorate for Higher Education and Skills (project No. UTF-2021/10210). A.K.-B. acknowledges the support from the Silesian University of Technology through a pro-quality grant, project number 14/030/SDU/10-27-01.

## Conflict of Interest

The authors declare no conflict of interest.

## Author Contributions

A.K. and A.G. conceptualized the work. A.A., J.H.P., D.W.J., H.L., W.J.L, R.H.H., R.Z., and Z.M. supplied a broad range of samples enabling this library-wide data collection. A.A. performed XRD characterization. J.G.F. and C.B. performed TEM analysis. J.Z. performed DFT calculations. A.G collected optical transmission, reflection, and emission data. A.K., A.C., A.K.B., and A.G. designed and performed nano-FTIR measurements. A.G. and A.K. prepared the draft of the paper and finalized it using proposals from all coauthors. A.K., Ø.P., W.J.L., Z.M. contributed to funding acquisition and administration of their parts of the activities. All coauthors have commented and agreed on publishing the finalized manuscript.

## Data Availability Statement

The data that support the findings of this study are available from the corresponding author upon reasonable request.

# Supplementary Material

## Optical library of $Ga_2O_3$ polymorphs

Augustinas Galeckas,[1,a)] Adrian Cernescu,[2] Anna Kaźmierczak-Bałata,[3] Javier García-Fernández,[1] Calliope Bazioti,[1] Alexander Azarov,[1] Ji-Hyeon Park,[4] Dae-Woo Jeon,[4] Halin Lee,[5] Won-Jae Lee,[5] Ray-Hua Horng,[6] Rui Zhu,[7,8] Zengxia Mei,[7,8] Øystein Prytz,[1] and Andrej Kuznetsov[1]

[1] University of Oslo, Department of Physics, Centre for Materials Science and Nanotechnology, PO Box 1048 Blindern, 0316 Oslo, Norway
[2] attocube systems AG, 85540 Haar, Germany
[3] Institute of Physics, Silesian University of Technology, 44-100 Gliwice, Poland
[4] Korea Institute of Ceramic Engineering & Technology, Jinju 52851, Republic of Korea
[5] Dong-Eui University, Department of Advanced Materials Engineering, 47340 Busan, Republic of Korea
[6] Institute of Electronics, National Yang Ming Chiao Tung University, Hsinchu 30010, Taiwan
[7] Songshan Lake Materials Laboratory, 523808 Dongguan, Guangdong, P. R. China
[8] Institute of Physics, Chinese Academy of Sciences, 100190 Beijing, P. R. China

Authors to whom correspondence should be addressed: [a)] augustinas.galeckas@fys.uio.no, [b)] andrej.kuznetsov@fys.uio.no

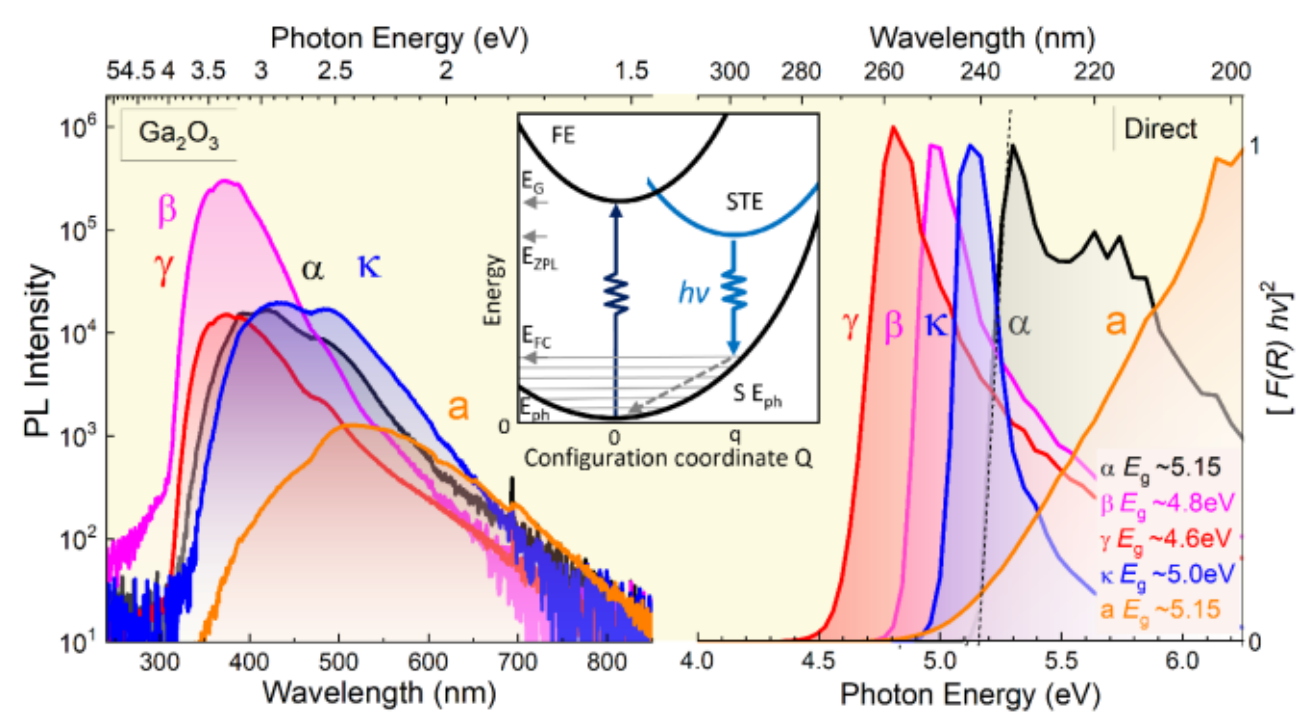

**Supplement #1: Absorption properties** (R%T% all samples)

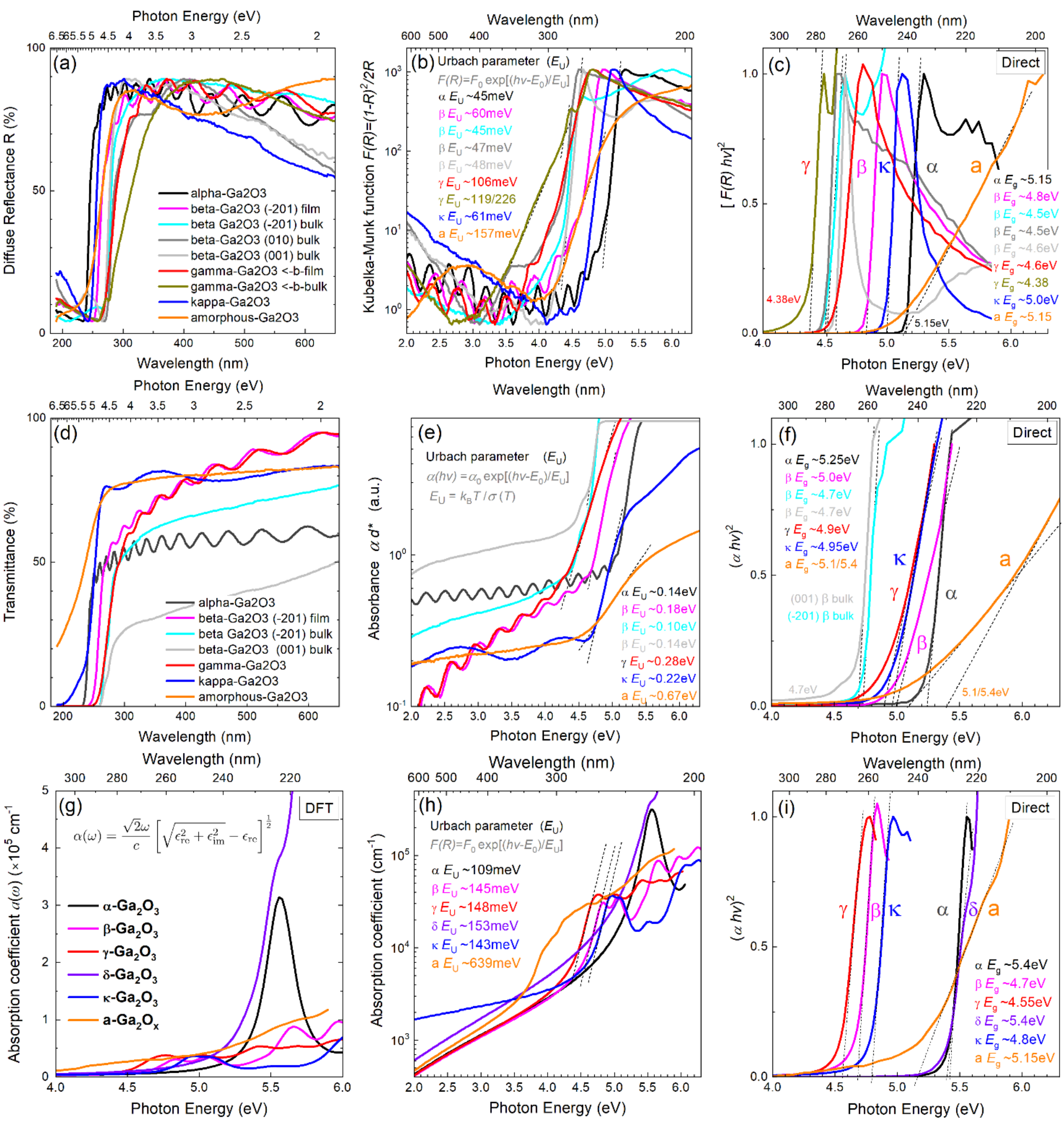

**Supplementary Figure S1**. Optical bandgap assessment of the *α*-, *β*-, *γ*-, *δ*-, and *κ*-$Ga_2O_3$ polymorphs, as well as amorphous a-$Ga_2O_x$, from room-temperature diffuse-reflectance and transmittance measurements: (a) Diffuse-reflectance spectra (normalized), (b) Kubelka-Munk function *F(R)* versus incident photon energy *hν*, (c) Tauc plot of the modified Kubelka-Munk function $[F(R)h\nu]^2$ assuming direct-allowed transitions. (d) Transmittance spectra (normalized), (e) Absorbance (*α d*) as a function of photon energy *hν*. (f) Tauc plot of the absorbance $[\alpha h\nu]^2$ considering direct-allowed transitions. The straight lines are linear extrapolations providing the Urbach energy $E_U$ (from the inverse slopes in panels b, e, h) and the bandgap $E_g$ parameters (from the energy-axis intercepts in panels c, f, i). For comparison with experiment, theoretical absorption spectra obtained from DFT calculations are processed in the same manner and shown in panels (g-i).

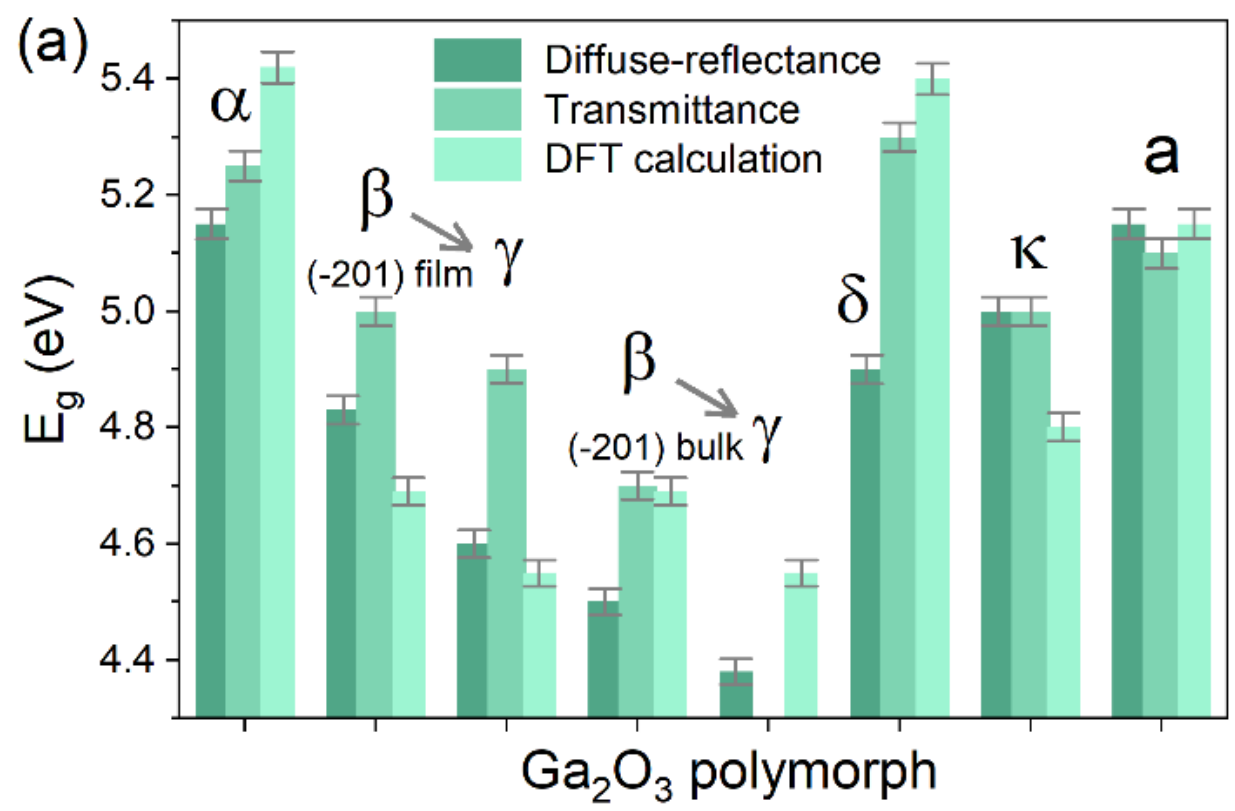

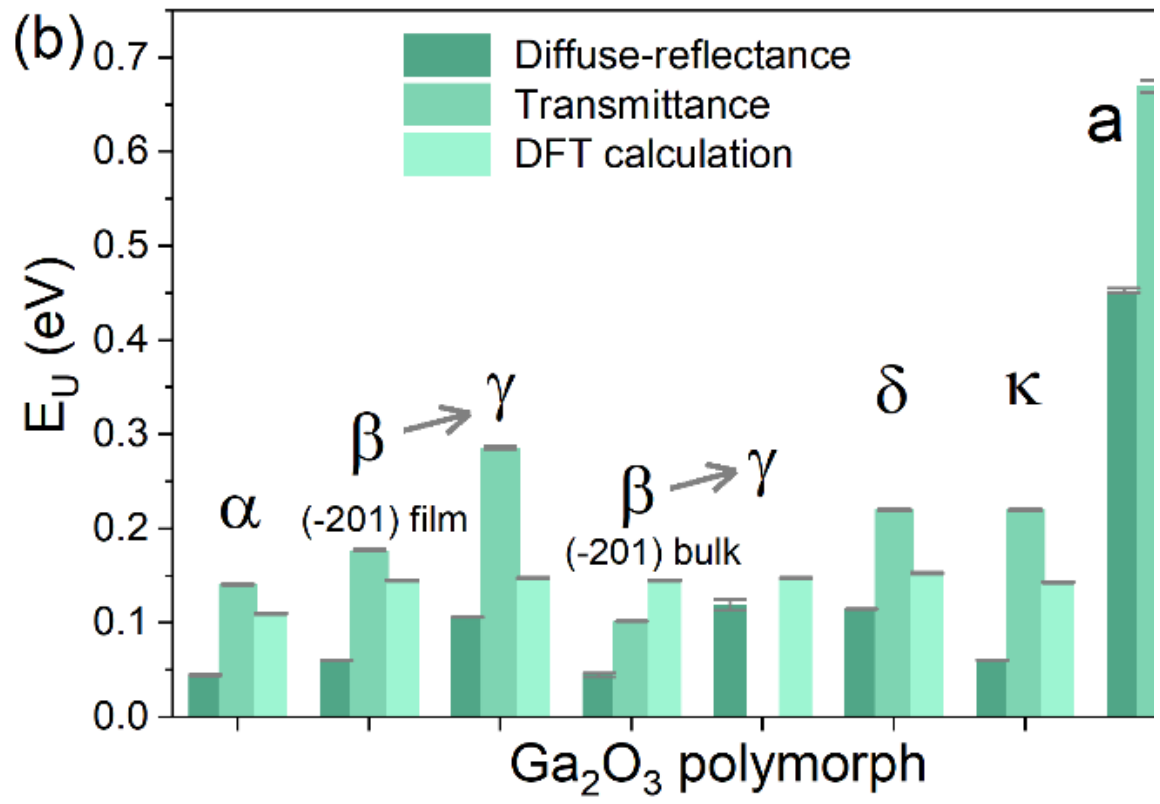

**Supplementary Figure S2**. Comparison of the (a) optical bandgap and (b) Urbach energy parameters for the *α*-, *β*-, *γ*-, *δ*-, and *κ*-$Ga_2O_3$ polymorphs, as well as amorphous a-$Ga_2O_x$, obtained from diffuse-reflectance and transmittance measurements, together with values derived from DFT-calculated absorption spectra.

**TABLE SI**. Summary of optical bandgap ($E_g$) and Urbach energy ($E_U$) parameters from diffuse-reflectance and transmittance measurements.

| Polymorph / structure | Diffuse Reflectance | | Transmittance | |
|---|---|---|---|---|
| | $E_g$ (eV) | $E_U$ (meV) | $E_g$ (eV) | $E_U$ (meV) |
| α-$Ga_2O_3$ film | 5.15 ± 0.03 | 45 ± 0.6 | 5.25 ± 0.03 | 141 ± 0.7 |
| β-$Ga_2O_3$ (-201) film | 4.83 ± 0.02 | 60 ± 0.3 | 5.00 ± 0.02 | 177 ± 0.9 |
| β-$Ga_2O_3$ (-201) bulk | 4.50 ± 0.02 | 45 ± 2 | 4.70 ± 0.02 | 102 ± 0.5 |
| β-$Ga_2O_3$ (010) bulk | 4.50 ± 0.02 | 47 ± 0.9 | | |
| β-$Ga_2O_3$ (001) bulk | 4.60 ± 0.02 | 48 ± 0.9 | 4.70 ± 0.02 | 137 ± 0.7 |
| γ-$Ga_2O_3$ film | 4.60 ± 0.02 | 106 ± 0.5 | 4.90 ± 0.02 | 286 ± 1.4 |
| γ /β-$Ga_2O_3$ bilayer | 4.38 ± 0.02 | 119 ± 0.6 | | |
| δ-$Ga_2O_3$ film | 4.75 ± 0.02 | 115 ± 0.5 | 4.90 ± 0.02 | 350 ± 1.4 |
| κ-$Ga_2O_3$ film | 5.00 ± 0.02 | 61 ± 0.3 | 5.00 ± 0.02 | 220 ± 1.1 |
| a-$Ga_2O_x$ film | 5.15 ± 0.02 | 157 ± 1.4 | 5.10 ± 0.03 | 670 ± 6.8 |

**Supplement #2: Emission properties** **(PL/ (010) (201) bulk)**

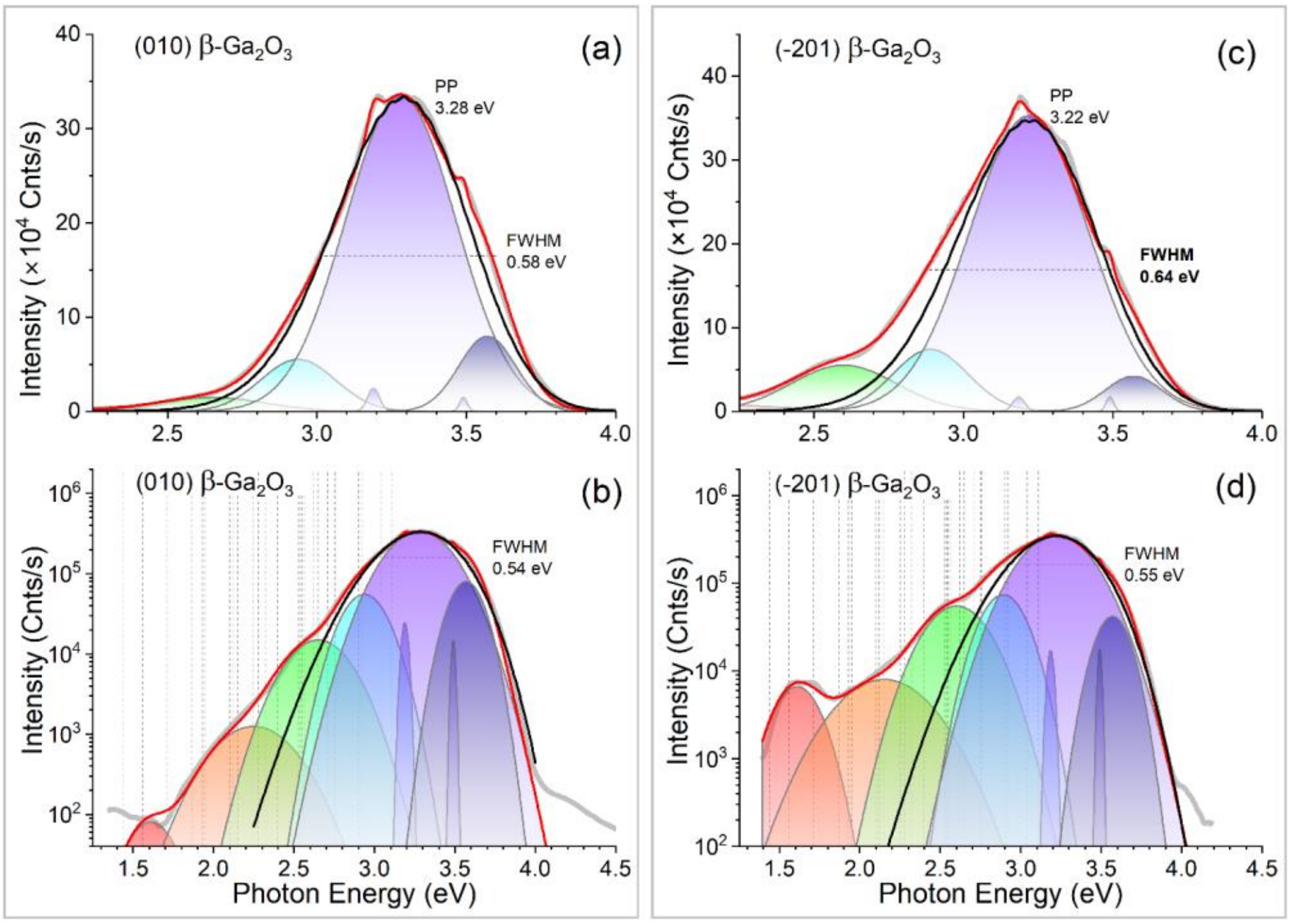

**Supplementary Figure S3.** PL spectra of monoclinic $\beta$-$Ga_2O_3$ polymorph obtained at 10K from (010) and (-201) oriented single-crystals plotted on linear and semi-log scales in panels (a,b) and (c,d), respectively. Gaussian deconvolution components are represented by grey curves. Vertical dashed lines are markers of the optical transitions (experimentally established or theoretically predicted) in the literature reports (see Table SII). Vertical dashed lines in panels (b,d) mark optical transitions that have been experimentally established or theoretically predicted in the literature reports [57-63] (see Table SII).

**Supplement #3: Near-field scanning nanoFTIR** (orientation)

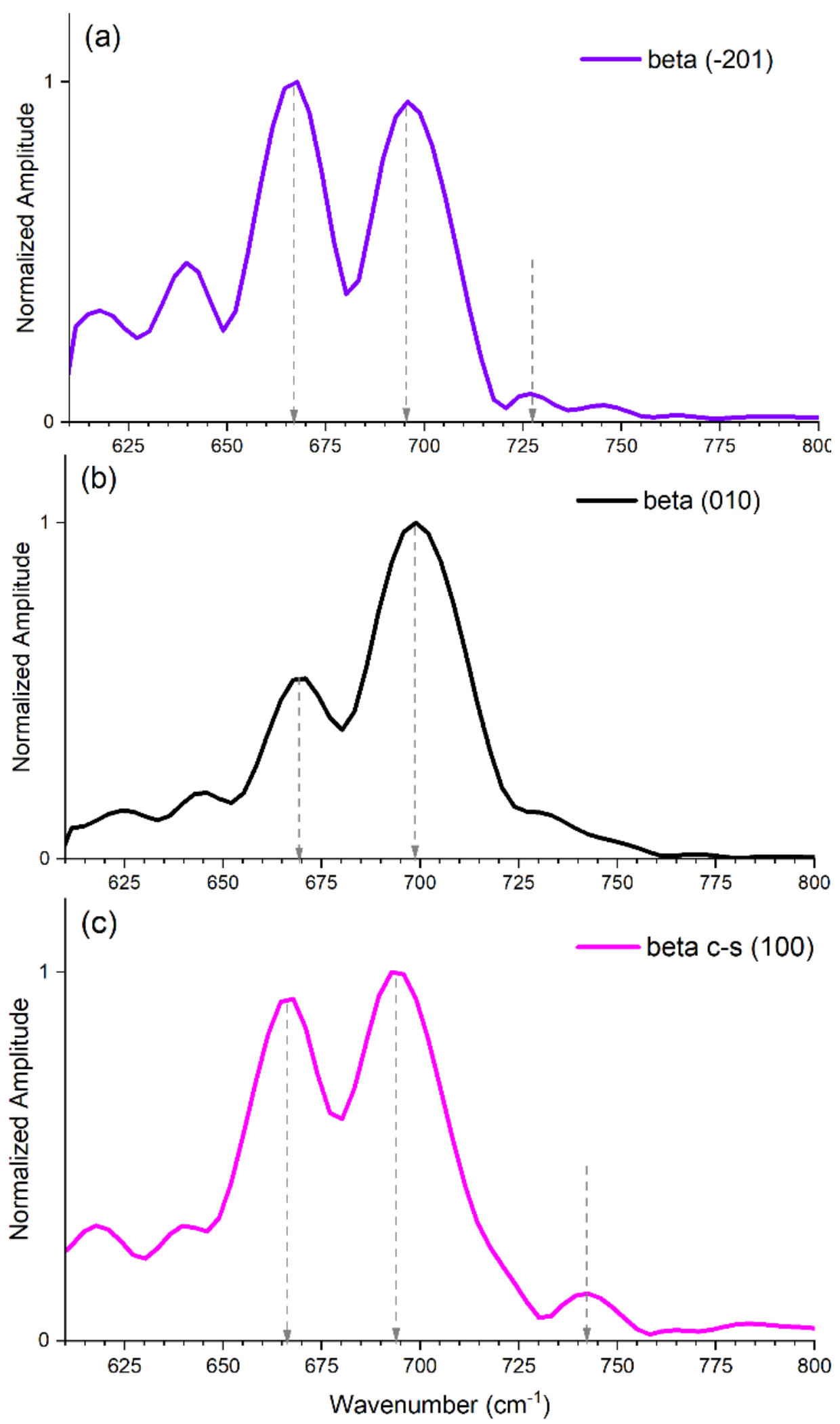

**Supplementary Figure S4.** Nano-FTIR spectra obtained at 300K for different crystal orientations of $\beta$-$Ga_2O_3$.

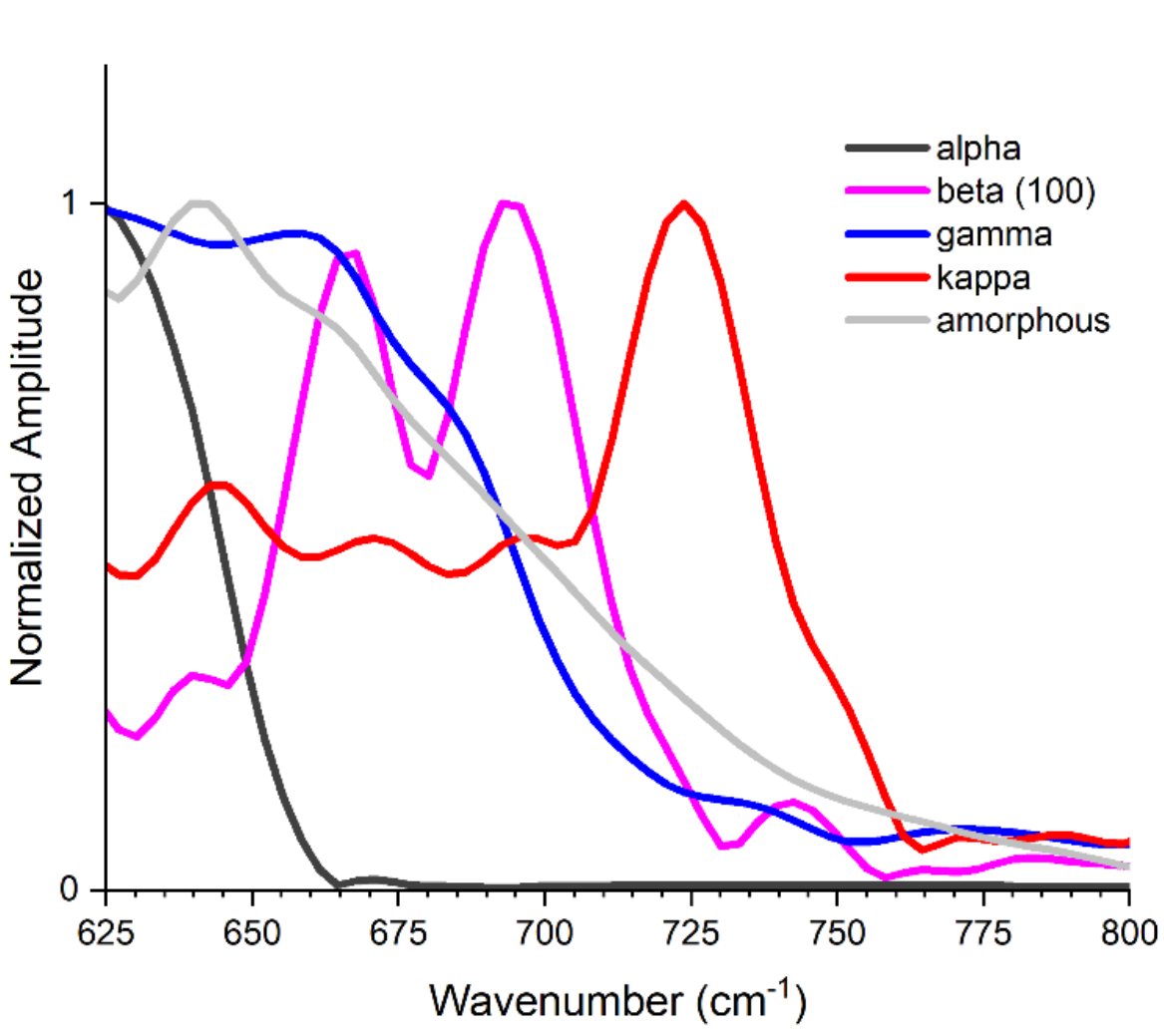

**Supplementary Figure S5.** Nano-FTIR spectra obtained at 300K for $\alpha$-, $\beta$-, and $\kappa$-$Ga_2O_3$ polymorphs, as well as amorphous a-$Ga_2O_X$.